\documentclass[a4paper,fleqn,usenatbib]{mnras}

\usepackage{newtxtext,newtxmath}

\usepackage[T1]{fontenc}
\usepackage{ae,aecompl}

\usepackage{siunitx}

\usepackage{subfigure}

\usepackage{graphicx}	
\usepackage{amsmath}	
\usepackage{amssymb}	

\setcitestyle{notesep={}}

\DeclareSIUnit \solarMass {\mbox{$M_\odot$}}
\DeclareSIUnit \year {yr}
\DeclareSIUnit \gauss {G}
\DeclareSIUnit \erg {erg}
\DeclareSIUnit \electronVolt {eV}

\newcommand \currsci {Curr. Sci}

\title[Population Synthesis of Accreting Neutron Stars]{Population Synthesis of Accreting Neutron Stars Emitting Gravitational Waves}

\author[F. Gittins and N. Andersson]{
Fabian Gittins$^{1,2}$\thanks{E-mail: f.w.r.gittins@soton.ac.uk}
and Nils Andersson$^{1,2}$
\\
$^{1}$STAG Research Centre, University of Southampton, Southampton, SO17 1BJ, UK\\
$^{2}$Mathematical Sciences, University of Southampton, Southampton, SO17 1BJ, UK
}

\date{Accepted XXX. Received YYY; in original form ZZZ}

\pubyear{2019}

\begin{document}
\label{firstpage}
\pagerange{\pageref{firstpage}--\pageref{lastpage}}
\maketitle

\begin{abstract}
The fastest-spinning neutron stars in low-mass X-ray binaries, despite having undergone millions of years of accretion, have been observed to spin well below the Keplerian break-up frequency. We simulate the spin evolution of synthetic populations of accreting neutron stars in order to assess whether gravitational waves can explain this behaviour and provide the distribution of spins that is observed. We model both persistent and transient accretion and consider two gravitational-wave-production mechanisms that could be present in these systems: thermal mountains and unstable \textit{r}-modes. We consider the case of no gravitational-wave emission and observe that this does not match well with observation. We find evidence for gravitational waves being able to provide the observed spin distribution; the most promising mechanisms being a permanent quadrupole, thermal mountains and unstable \textit{r}-modes. However, based on the resultant distributions alone it is difficult to distinguish between the competing mechanisms.
\end{abstract}

\begin{keywords}
accretion, accretion discs -- gravitational waves -- stars: neutron -- stars: rotation -- X-rays: binaries
\end{keywords}



\section{Introduction}

Neutron stars (NSs) are among the most compact objects that we observe in our universe. For this reason, they are able to spin up to extremely high frequencies. The classical picture for the evolution of rapidly-spinning NSs begins with a NS accreting gas via a circumstellar accretion disc from their companion in a low-mass X-ray binary \citep[LMXB;][]{1982Natur.300..728A, 1982CSci...51.1096R}. This process causes the NS to spin up and, eventually, the NS accretes all the gas from its companion such that all that is left of the binary is a radio millisecond pulsar (RMSP). This scenario should, in theory, have no difficulty in spinning NSs up to their centrifugal break-up frequency \citep{1994ApJ...424..823C}, which is generally above $\sim \SI{1}{\kilo\hertz}$ for most equations of state \citep{2007PhR...442..109L}. However, the fastest-spinning pulsar that has been observed to date is PSR J1748-2446ad which spins at \SI{716}{\hertz} \citep{2006Sci...311.1901H}; well below the limit set by the break-up frequency. In fact, the distribution of spins for both LMXBs and RMSPs has been shown to have a statistically significant cut-off at \SI{730}{\hertz} \citep{2003Natur.424...42C, 2010ApJ...722..909P}.

Accreting millisecond X-ray pulsars (AMXPs) are a sub-class of LMXBs which have been spun up to millisecond periods through accretion \citep[see][ for a review on AMXPs]{2012arXiv1206.2727P}. They are characterised by accretion rates $\gtrsim \SI{e-11}{\solarMass\per\year}$ and comparatively weak magnetic fields ($\sim \SI{e8}{\gauss}$). Another important sub-class of LMXBs are the nuclear-powered X-ray pulsars (NXPs). These pulsars show short-lived burst oscillations during thermonuclear burning on their surfaces and are distinct from AMXPs due to being powered by nuclear burning rather than accretion. 19 AMXPs and 11 NXPs have been observed to date.

\citet{2018A&A...620A..69H} show that the observed rotation rate limit on NSs does not correspond to centrifugal break-up and argue that additional spin-down torques are required to explain this effect. It is unclear what physical process prevents these NSs from spinning up to sub-millisecond periods. One candidate is the interaction between the magnetic field and the accretion disc \citep{1978ApJ...223L..83G, 1997ApJ...490L..87W, 2005MNRAS.361.1153A}. \citet{2012ApJ...746....9P} demonstrated that a magnetic-field strength at the magnetosphere of $\sim \SI{e8}{\gauss}$ could be enough to explain the deficiency in accreting NSs above $\sim \SI{700}{\hertz}$. More recently, it has been shown that transient accretion can have a significant impact on the spin evolution of an accreting NS \citep{2017MNRAS.470.3316D, 2017ApJ...835....4B}. In fact, \citet{2017ApJ...835....4B} noted that in the case of transient accretion, a magnetosphere of $\sim \SI{e8}{\gauss}$ would no longer be sufficient to explain the spin limit. It was suggested by \citet{1998ApJ...501L..89B} and \citet{1999ApJ...516..307A} that one would observe a spin-frequency limit to accreting NSs if they were emitting gravitational waves (GWs), thus providing a torque to balance the accretion torques \citep{1978MNRAS.184..501P, 1984ApJ...278..345W}.

Rapidly-rotating NSs have been the subject of many GW searches. These include searches of known radio pulsars \citep{2004PhRvD..69h2004A, 2005PhRvL..94r1103A, 2007PhRvD..76d2001A, 2008ApJ...683L..45A, 2010ApJ...713..671A, 2017PhRvD..96l2006A, 2017ApJ...839...12A, 2018PhRvL.120c1104A, 2011PhRvD..83d2001A, 2011ApJ...737...93A, 2014ApJ...785..119A, 2015PhRvD..91b2004A}, along with wide-parameter surveys for unknown pulsars \citep{2005PhRvD..72j2004A, 2007PhRvD..76h2001A, 2008PhRvD..77b2001A, 2009PhRvD..79b2001A, 2016PhRvD..94d2002A, 2017PhRvD..96f2002A, 2018PhRvD..97j2003A, 2012PhRvD..85b2001A, 2013PhRvD..87d2001A}. Assuming torque balance, then the brightest X-ray sources should thus be the loudest in GW emission. Hence, Scorpius X-1 has long been considered a potential candidate for GW emission and has been the focus of a number of targeted GW searches \citep{2007PhRvD..76h2001A, 2017PhRvD..95l2003A, 2017ApJ...847...47A, 2015PhRvD..91f2008A}. These searches, while only yielding upper limits so far, have helped develop and improve data analysis techniques that can be used in the future, when GW detectors further increase their sensitivities. Detecting GWs from rotating NSs will be a challenge, since only a few of the most rapidly-accreting NSs are detectable with the current generation of GW detectors \citep{2008MNRAS.389..839W}. The main limiting factors in detecting GWs from these systems are the precision with which the spin and the orbital parameters are measured. If these are not well known, then it becomes computationally very expensive to run GW searches, since the searches need to run over a large parameter space.

\citet{2017ApJ...850..106P} performed a statistical analysis of accreting NSs in LMXBs and found that there is statistical evidence for the distribution to comprise two sub-populations: one at relatively low spin frequencies with an average of $\approx \SI{300}{\hertz}$ and the other at higher frequencies with an average of $\approx \SI{575}{\hertz}$. The two sub-populations were shown to transition from one to the other in the region around $\approx \SI{540}{\hertz}$. \citet{2017ApJ...850..106P} suggested that GWs would provide a physical meaning to this transition region -- above this point, GWs become significant. However, \citet{2017ApJ...850..106P} noted that some of the sources that belong to the fast sub-population have a spin behaviour that is not straightforwardly reconciled with GW scenarios. \citep[See][ for additional work on the spin-frequency distribution of millisecond pulsars.]{2008AIPC.1068..130H, 2014A&A...566A..64P}

In this paper, we investigate whether an additional component is needed in order to explain the spin evolution of accreting NSs. We consider GW emission as the source of such a component and explore what might be the dominant GW-production mechanism. This paper is structured as follows. To begin with, in Section~\ref{sec:AccretionLMXBs}, we introduce the basic theory regarding accretion in LMXBs and discuss how transient accretion can affect the spin evolution of accreting NSs. In Section~\ref{sec:GWs}, we provide a brief review of gravitational radiation in these systems and the different mechanisms that can give rise to such radiation. We describe our model for the spin evolution of accreting NSs in Section~\ref{sec:SpinModel}. In Section~\ref{sec:SimulatedPopulations}, we summarise the results of our NS population simulations, which include the different GW-production mechanisms. Finally, we conclude and suggest future work in Section~\ref{sec:Conc}.

\section{Accretion in Low-Mass X-ray Binaries}
\label{sec:AccretionLMXBs}

In a LMXB, the companion star has overfilled its Roche lobe and is donating matter to the NS through the inner Lagrange point. Given this donated gas has some large specific angular momentum it cannot be transferred directly to the surface of the NS and instead forms a circumstellar accretion disc around it. The gas from the accretion disc is channelled onto the magnetic poles of the NS along the magnetic-field lines. This channelling occurs at what is known as the \textit{magnetospheric radius}, $r_\text{m}$, which is the characteristic radius where the magnetic field dominates interactions. At this boundary the magnetic field can, in turn, be distorted by the accreting gas. This coupling, between the field lines and the disc, results in a torque acting on the star that can spin it up or down depending on the relative difference between $r_\text{m}$ and the \textit{co-rotation radius},
\begin{equation}
	r_\text{c} \equiv \left( \frac{G M}{\Omega^2} \right)^{1 / 3},
\end{equation}
which is the location of a Keplerian disc that rotates with the same frequency of the NS, where $G$ is the gravitational constant, $M$ is the mass of the NS, $\Omega = 2 \pi \nu = 2 \pi / P$ is its angular frequency, $\nu$ is the spin frequency and $P$ is the spin period. If $r_\text{m} < r_\text{c}$, the NS spins slower than the accretion disc and so the gas that is channelled onto the NS has greater specific angular momentum than it, thus acting to spin it up. Conversely, if $r_\text{m} > r_\text{c}$, the NS spins faster than the disc, which spins it down.

The magnetospheric radius is a somewhat poorly understood quantity. It is generally defined as the point where the kinetic energy of in-falling gas becomes comparable to the magnetic energy of the magnetosphere. For the straightforward case where the gas is radially accreted onto the NS, one can calculate the Alfv{\'e}n radius, $r_\text{A}$, from:
\begin{equation}
	\frac{1}{2} \rho(r_\text{A}) v(r_\text{A})^2 = \frac{B(r_\text{A})^2}{8 \pi},
\end{equation}
where $\rho(r_\text{A})$, $v(r_\text{A})$ and $B(r_\text{A})$ are the gas density, gas velocity and magnetic-field strength, respectively, at $r_\text{A}$. This calculation gives the standard expression for the magnetospheric radius \citep{1972A&A....21....1P}:
\begin{equation}
	r_\text{A} = \left( \frac{\mu^4}{2 G M \dot{M}^2} \right)^{1/7},
\end{equation}
where $\mu = B R^3$ is the magnetic moment of the NS, $\dot{M}$ is the mass-accretion rate from the disc to the NS surface and $R$ is the radius of the NS. This picture becomes more complicated when considering accretion from a circumstellar disc. A factor $\xi$ of order unity is introduced to correct for the non-spherical geometry of the problem and also account for the extended transition region between where mass is accreted onto the star and where mass is ejected in an outflow. This gives the magnetospheric radius as
\begin{equation}
	r_\text{m} = \xi r_\text{A} = \xi \left( \frac{\mu^4}{2 G M \dot{M}^2} \right)^{1/7}.
\end{equation}
Typically, $\xi$ is assumed to fall in the range $0.5 - 1.4$ \citep{1996ApJ...465L.111W}. This correction demonstrates that $r_\text{m}$ is sensitive to the coupling between the field lines and the accretion disc which plays a key role in understanding the magnetospheric radius.

\subsection{Accretion-Torque Models}
\label{sec:AccretionTorque}

The spin evolution of a NS is dictated by the torques acting upon the star. By measuring the time derivative of the spin period, $\dot{P}$, one can gain insight into the physics of accretion processes, as well as other aspects such as magnetic-field strengths and GW emission. The variation in spin is related to the torque exerted onto the NS by the standard expression
\begin{equation}
	N = -\frac{2 \pi I \dot{P}}{P^2}, 
	\label{eq:TorqueSpinDerivative}
\end{equation}
where $I$ is the stellar moment of inertia.

For a NS accreting from an accretion disc truncated at the magnetospheric transition, with $r_\text{m} < r_\text{c}$, the standard torque is (neglecting magnetic-field effects)
\begin{equation}
	N_\text{acc} = \dot{M} r_\text{m}^2 \Omega_\text{K}(r_\text{m}) = \dot{M} \sqrt{G M r_\text{m}},
	\label{eq:AccretionDiscN}
\end{equation}
where $\Omega_\text{K}(r_\text{m})$ is the Keplerian angular velocity at $r_\text{m}$. For $r_\text{m} > r_\text{c}$, the coupling between the magnetosphere and the accretion disc becomes important as the magnetic-field lines are threaded through the disc so extra torques due to magnetic stresses come into play. \citet{1979ApJ...234..296G} developed an accretion model based on detailed calculations of this coupling. The torque from this model predicts
\begin{equation}
	\dot{P} \approx \num{-5.0e-5} M_{1.4}^{3/7} I_{45}^{-1} \mu_{30}^{2/7} \left[ \left( \frac{P}{\SI{1}{\second}} \right) \dot{M}_{-9}^{3/7} \right]^2 n(\omega_\text{s}) \, \si{\second\per\year},
	\label{eq:GhoshLambPdot}
\end{equation}
where $M_{1.4} = M / \SI{1.4}{\solarMass}$, $R_6 = R / \SI{10}{\kilo\metre}$, $I_{45} = I / \SI{e45}{\gram\centi\metre\squared}$, $\mu_{30} = \mu / \SI{e30}{\gauss\centi\metre\cubed}$, $\dot{M}_{-9} = \dot{M} / \SI{e-9}{\solarMass\per\year}$ and $n(\omega_\text{s})$ is the dimensionless torque which accounts for the magnetic field-accretion disc coupling and is a function of the fastness parameter $\omega_\text{s}$. The fastness parameter is defined as the ratio of the NS spin frequency to the Keplerian orbital frequency at the magnetospheric boundary,
\begin{equation}
	\omega_\text{s} \equiv \frac{\Omega}{\Omega_\text{K}(r_\text{m})} \approx \num{3.1} \, \xi^{3/2} M_{1.4}^{-5/7} \mu_{30}^{6/7} \left[ \left( \frac{P}{\SI{1}{\second}} \right) \dot{M}_{-9}^{3/7} \right]^{-1}.
\end{equation}
The sign of $n(\omega_\text{s})$ depends on whether the NS accretes the gas and spins up (the `slow rotator' regime, $\omega_\text{s} < 1$) or ejects the gas and spins down \citep[the `fast rotator' regime, $\omega_\text{s} > 1$;][]{1995ApJ...449L.153W}. It is interesting to note that a NS can still be spun down at long spin periods ($P \gg \SI{1}{\second}$) as the magnetic field can be strong enough to mean it would still be classified as a fast rotator.

\citet{2014MNRAS.437.3664H} introduced a simple approximation to the \citeauthor{1979ApJ...234..296G} model by considering angular momentum changes on the NS. Matter accreting at the magnetosphere, $r_\text{m}$, has specific angular momentum
\begin{equation}
	l_\text{acc} = \pm r_\text{m}^2 \Omega_\text{K}(r_\text{m}),
\end{equation}
where the sign of $l_\text{acc}$ depends on whether there is prograde rotation between the accretion disc and the NS ($l_\text{acc} > 0$) or retrograde rotation ($l_\text{acc} < 0$). Prograde rotation will be assumed. What must also be accounted for is matter that is ejected from the NS, which will carry specific angular momentum
\begin{equation}
	l_\text{m} = r_\text{m}^2 \Omega.
\end{equation}
Both of these effects produce a torque on the NS. The net torque is obtained by summing these contributions:
\begin{equation}
	N = \dot{M} (l_\text{acc} - l_\text{m}) = \dot{M} r_\text{m}^2 \Omega_\text{K}(r_\text{m}) (1 - \omega_\text{s}).
	\label{eq:HoN}
\end{equation}
In this relation, one can see the standard spin-up torque due to disc accretion (Equation~\ref{eq:AccretionDiscN}) which is corrected by the fastness parameter to account for interactions spinning down the NS. (This model phenomenologically accounts for effects such as accretion disc-magnetic field coupling and outflows.) This expression can be related to the change in spin period using Equation~(\ref{eq:TorqueSpinDerivative}) to obtain
\begin{equation}
	\dot{P} \approx \num{-8.1e-5} \, \xi^{1/2} M_{1.4}^{3/7} I_{45}^{-1} \mu_{30}^{2/7} \left[ \left( \frac{P}{\SI{1}{\second}} \right) \dot{M}_{-9}^{3/7} \right]^2 (1 - \omega_\text{s}) \, \si{\second\per\year}.
	\label{eq:HoPdot}
\end{equation}
It is clear from Equation~(\ref{eq:HoPdot}) that the fastness parameter dictates whether the NS spins up or down.

A commonly considered aspect of a NS's spin evolution is the spin equilibrium. This occurs when the spin rate is gradually adjusted until the net torque on the star is approximately zero and the accretion flow is truncated at the magnetospheric radius, $r_\text{m} \simeq r_\text{c}$. When a NS reaches spin equilibrium, it is straightforward to estimate its magnetic field, assuming that the accretion rate and spin are known. One can estimate the spin-equilibrium period, $P_\text{eq}$, from Equation~(\ref{eq:HoPdot}) by setting $\dot{P} = 0$, when $\omega_\text{s} = 1$:
\begin{equation}
	P_\text{eq} \approx \num{8.2} \, \xi^{3/2} M_{1.4}^{-5/7} \left(\frac{\mu}{\SI{e26}{\gauss\centi\metre\cubed}}\right)^{6/7} \left(\frac{\dot{M}}{\SI{e-11}{\solarMass\per\year}}\right)^{-3/7} \, \si{\milli\second},
\end{equation}
where we have scaled the period to characteristic AMXP values.

The magnetic-field lines rotate with the NS. This produces magnetic-dipole radiation which causes the NS to spin down. The torque due to this is described by
\begin{equation}
	N_\text{EM} = - \frac{2 \mu^2 \Omega^3}{3 c^3},
\end{equation}
where $c$ is the speed of light in a vacuum. The change in spin due to magnetic-dipole radiation is 
\begin{equation}
	\dot{P}_\text{EM} \approx \num{3.1e-8} \mu_{30}^2 I_{45}^{-1} \left( \frac{P}{\SI{1}{\second}} \right)^{-1} \si{\second\per\year}.
	\label{eq:EMPdot}
\end{equation}
The vast majority of pulsars are isolated and their spin evolution can be generally described by magnetic-dipole radiation. However, in the case of rapidly-accreting NSs this effect can be essentially negligible. There are more accurate numerical models one can use to describe these torques \citep[e.g., see][]{2006ApJ...648L..51S}.

\subsection{Transient Accretion}
\label{sec:TransientAccretion}

Up until now, the accretion rate has been implicitly assumed to be steady. However, many LMXBs exhibit long periods of quiescence, which can be of the order of months to years, and short transient outbursts, which can last from days to weeks. These outbursts are believed to be caused by instabilities in the accretion disc and occur when the mass-accretion rate rises above a certain threshold \citep[see, e.g.,][]{1997ASPC..121..351L}. As the companion star donates a steady flow of gas to the accretion disc, the disc gets larger and eventually reaches a critical mass to trigger an instability. This causes the accretion rate from the disc to the surface of the NS to increase by several orders of magnitude, giving rise to a transient outburst. Once the accretion disc has donated a sufficient amount of gas the system returns to a quiescent state until a new outburst occurs when the disc has accumulated enough mass from the companion and the cycle repeats \citep{2007A&ARv..15....1D}.

\citet{2017MNRAS.470.3316D} and \citet{2017ApJ...835....4B} have shown that transient accretion with a varying accretion rate has a significant impact on the spin evolution of a NS. Both found that for a given long-term average accretion rate, these transients can spin up NSs to rates several times higher than that of persistent accretors, however, it takes approximately an order of magnitude longer to reach these spin-equilibrium periods. This demonstrates that for transient systems, like most LMXBs, it is not accurate to assume a time-averaged accretion rate but instead one must consider the outburst/quiescence phases. \citet{2017MNRAS.470.3316D} noted that the two key changes when considering transient accretion are: that the torque over an outburst is significantly smaller than for the persistent case at a given accretion rate and that the equilibrium accretion rate is shifted to a lower value. This has the combined effect to increase the time it takes for a transient source to reach spin equilibrium and decrease its spin-equilibrium period.

\citet{2017MNRAS.470.3316D} and \citet{2017ApJ...835....4B} found that the spin-equilibrium period and time to reach spin equilibrium are sensitive to the features of the accretion profile. They found that by increasing the duration of an outburst by a factor of $10$ the spin-equilibrium period can decrease by up to a factor of $2$.

\begin{figure}
	\includegraphics[width=\columnwidth]{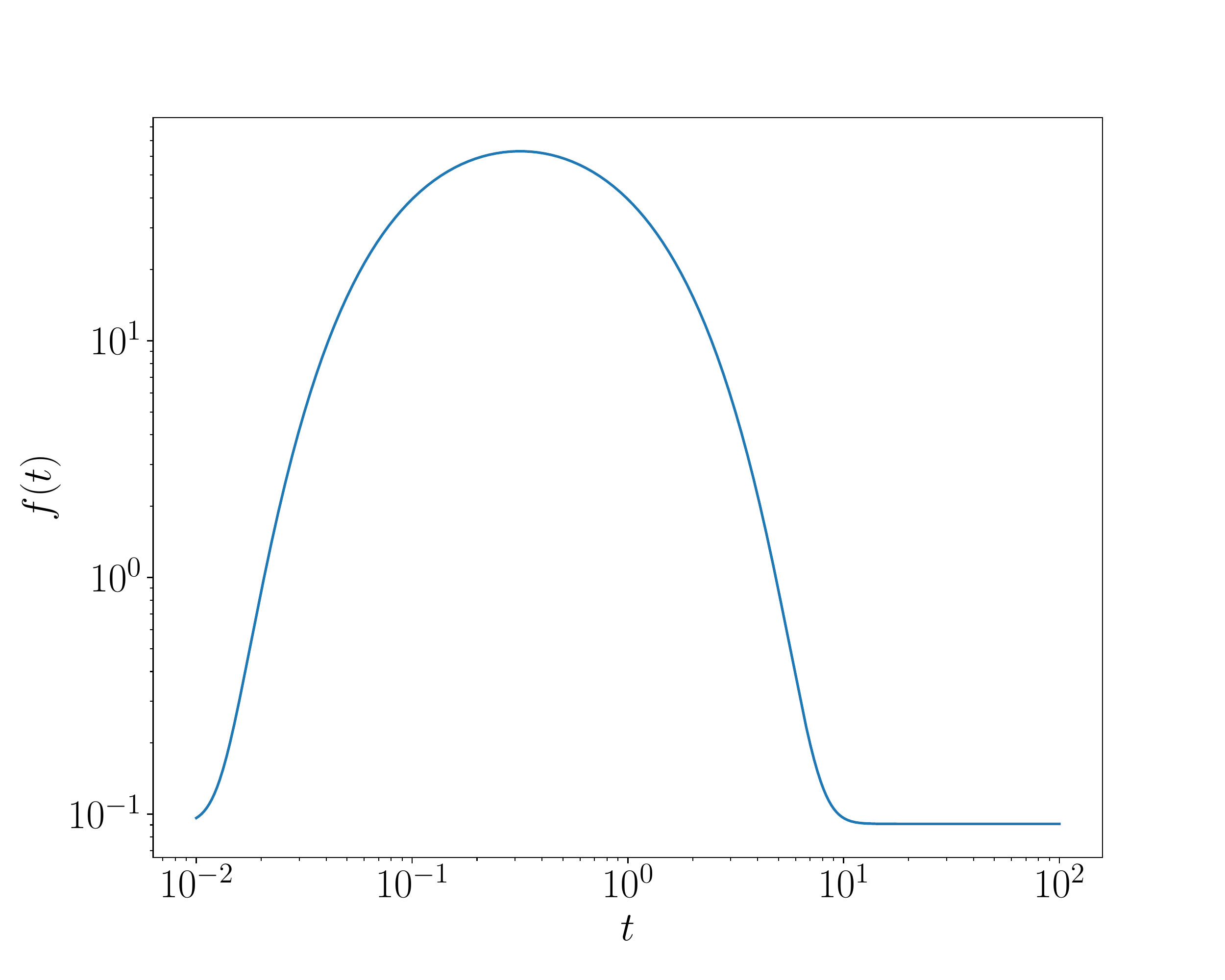}
	\caption{Accretion outburst profile, $f(t)$, as a function of time, $t$, where $F_\text{t} = 10$, $f_\text{max} / f_\text{min} = 693.97$ and $T_\text{recurrence} = 100$. The accretion rate and time have arbitrary units.}
	\label{fig:AccretionProfile}
\end{figure}

For their analysis, \citet{2017MNRAS.470.3316D} used a fast-rise, exponential-decay function to model the accretion profile \citep[whereas][ used a simple sawtooth function]{2017ApJ...835....4B}:
\begin{equation}
	f(t) = \exp \left( \sqrt{\frac{2}{F_\text{t}}} \right) \exp \left( - \frac{1}{10 t} - \frac{10 t}{F_\text{t}} \right) + f_\text{min},
	\label{eq:AccretionProfile}
\end{equation}
where $t$ denotes the time from the beginning of the outburst, $F_\text{t}$ is an approximate measure of the duration of the outburst and $f_\text{min}$ is the minimum. It should be noted that this function models a single outburst/quiescence cycle, so in order to model multiple cycles one repeats this after a given recurrence time, $T_\text{recurrence}$. Time has arbitrary units in this model. The ratio of the maximum to the minimum is
\begin{equation}
	\frac{f_\text{max}}{f_\text{min}} = \frac{1}{f_\text{min}} \exp \left( \frac{\sqrt{2} - 2}{\sqrt{F_\text{t}}} \right) + 1.
	\label{eq:AccretionRatio}
\end{equation}
This accretion profile requires two normalisations. The first normalisation chooses $f_\text{max} / f_\text{min}$ to obtain $f_\text{min}$ for a fixed $F_\text{t}$ using Equation~(\ref{eq:AccretionRatio}). The second normalisation is to demand that $\langle f(t) \rangle = 1$. This normalisation depends on $T_\text{recurrence}$ and results in $f_\text{min}$ no longer corresponding precisely to the minimum value. These normalisations allow one to choose the magnitude of the accretion outburst, with respect to the quiescent accretion rate, and also mean that one can simply choose an average accretion rate over one cycle by multiplying Equation~(\ref{eq:AccretionProfile}) by the chosen average. Thus, the time-dependant accretion rate is given by
\begin{equation}
	\dot{M}(t) = \langle \dot{M} \rangle f(t),
	\label{eq:AccretionRate}
\end{equation}
where $f(t)$ has been appropriately normalised. The canonical profile used by \citet{2017MNRAS.470.3316D} had $F_\text{t} = 10$, $f_\text{max} / f_\text{min} = 693.97$ and $T_\text{recurrence} = 100$ and is shown in Figure~\ref{fig:AccretionProfile}.

\section{Gravitational Radiation from Accreting Neutron Stars}
\label{sec:GWs}

Rotating NSs emit GWs if they are asymmetric about the axis of rotation. Such mass asymmetries are referred to as \textit{mountains}. For a spinning NS with a mass-quadrupole moment $Q_{2 2}$ the braking torque due to GWs is given by
\begin{equation}
	N_\text{GW} = - \frac{256 \pi}{75} \frac{G \Omega^5 Q_{2 2}^2}{c^5}.
	\label{eq:GWN}
\end{equation}
This corresponds to a spin-down rate of
\begin{equation}
	\dot{P}_\text{GW} \approx \num{1.4e-19} I_{45}^{-1} \left( \frac{Q_{2 2}}{\SI{e37}{\gram\centi\metre\squared}} \right)^2 \left( \frac{P}{\SI{1}{\second}} \right)^{-3} \, \si{\second\per\year}.
	\label{eq:GWPdot}
\end{equation}

In order to estimate how strong a quadrupole is needed in order to considerably influence the spin evolution of the NS it is useful to balance Equation~(\ref{eq:GWN}) with the accretion-magnetosphere torque from Equation~(\ref{eq:HoN}). This leads to
\begin{equation}
\begin{split}
	Q_{2 2} &\approx \num{4.2e37} \xi^{1 / 4} M_{1.4}^{3 / 14} \mu_{30}^{1 / 7} \dot{M}_{-9}^{3 / 7} \left(\frac{\nu}{\SI{500}{\hertz}}\right)^{- 5 / 2} \\
	&\quad \times (1 - \omega_\text{s}) \, \si{\gram\centi\metre\squared}.
	\label{eq:BalanceQ}
\end{split}
\end{equation}
For a typical AMXP with $B \sim \SI{e8}{\gauss}$ and $\dot{M} \sim \SI{e-11}{\solarMass\per\year}$, this gives a quadrupole moment of $Q_{2 2} \sim \SI{e36}{\gram\centi\metre\squared}$ in order to achieve spin equilibrium at $\nu \sim \SI{500}{\hertz}$. One can express a mass-quadrupole in terms of a moment of inertia ellipticity, defined as \citep[see, e.g.,][]{2005PhRvL..95u1101O}
\begin{equation}
	\epsilon \equiv \sqrt{\frac{8 \pi}{15}} \frac{Q_{2 2}}{I}.
\end{equation}
Therefore, in order to balance the accretion torque with GW spin-down, one requires $\epsilon \sim \num{e-9}$. This is far smaller than the maximum deformation a NS crust can sustain for most reasonable equations of state \citep{2013PhRvD..88d4004J}. Recent population-based analysis has suggested that $\epsilon \approx \num{e-9}$ is the minimum ellipticity of millisecond pulsars \citep{2018ApJ...863L..40W}.

\begin{figure}
	\includegraphics[width=\columnwidth]{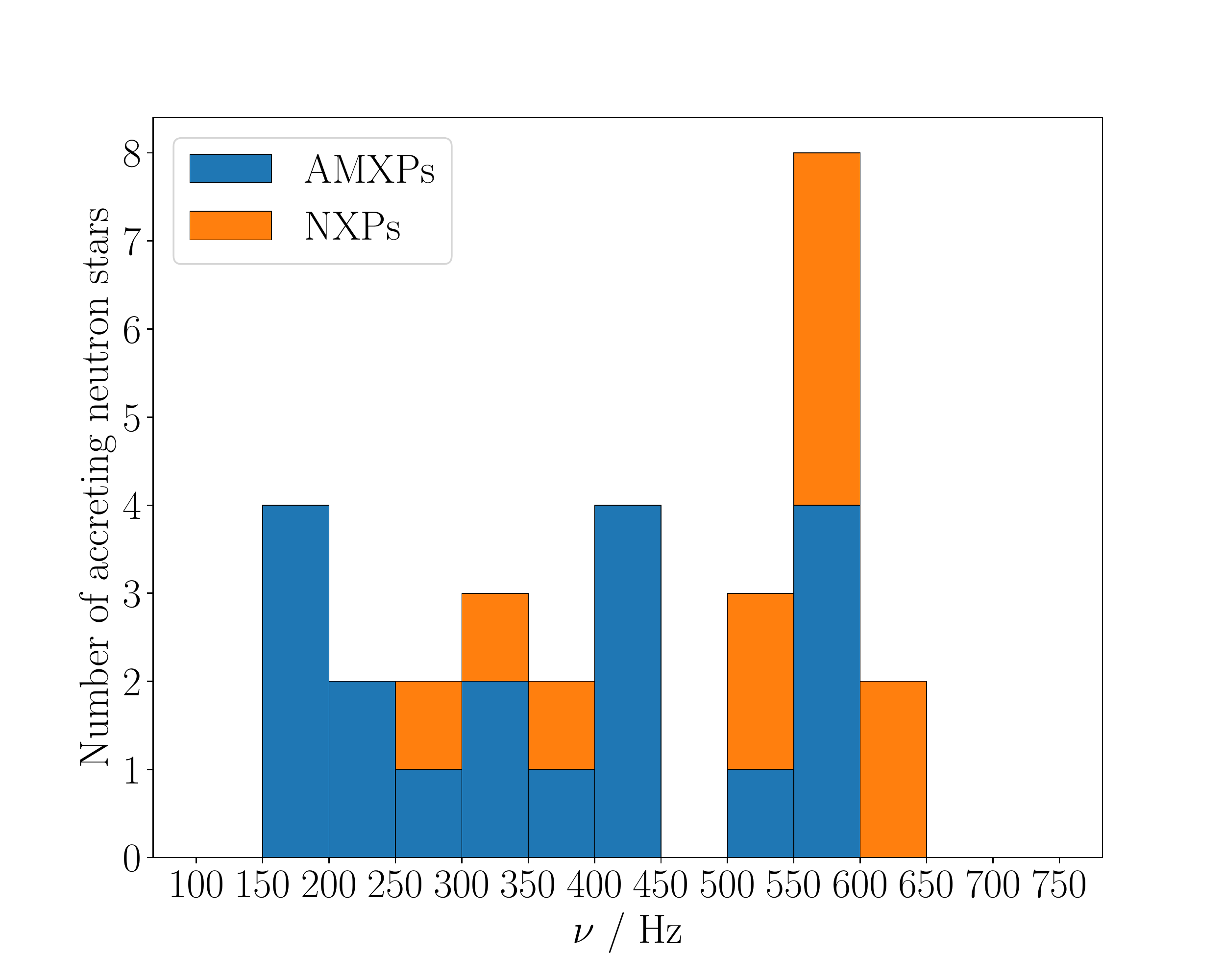}
	\caption{Distribution of spin frequencies for accreting NSs with millisecond periods. The accreting NS population comprises AMXPs, coloured in blue, and NXPs, coloured in orange.}
	\label{fig:SpinDistribution}
\end{figure}

An outstanding problem in understanding rapidly-spinning accreting NSs is their peculiar spin distribution (see Figure~\ref{fig:SpinDistribution}). It is this unusual shape and, in particular, the sharp cut-off at $\sim \SI{600}{\hertz}$ that has motivated the search for GWs from these systems. This is an appealing explanation since the braking torque due to GWs scales as the fifth power of the spin frequency for deformed, rotating NSs. \citet{2017ApJ...850..106P} have shown that among AMXPs and NXPs, there appear to be two sub-populations. One sub-population is at a relatively low spin-frequency, with a mean spin of $\approx \SI{300}{\hertz}$. The second sub-population has a higher peak and a mean of $\approx \SI{575}{\hertz}$. This faster sub-population has a very narrow range and is composed of a mixture of AMXPs and NXPs. The two sub-populations are separated by a transition region around $\approx \SI{540}{\hertz}$. \citet{2017ApJ...850..106P} argued that, when considering various accretion torque models, no model naturally explains the presence of a fast sub-population and postulated that whatever mechanism that causes this clustering it must set in quickly -- as soon as the pulsars reach a certain spin threshold. It was noted by \citet{2017ApJ...850..106P} that this is a subtly different problem to the one of accreting NSs not spinning close to their break-up frequency. These two problems make GWs a promising avenue to explore. GWs can help justify the transition region between the two sub-populations and provide a physical meaning to it (the region in which GW emission starts to become significant), and naturally explain the cut-off at $\sim \SI{600}{\hertz}$.

There are a number of different ways a mass asymmetry could arise in an accreting NS. \citet{1998ApJ...501L..89B} originally proposed that interior temperature asymmetries misaligned with respect to the spin-axis of the NS could produce a significant quadrupole through temperature-sensitive electron captures. Hotter regions of the crust would have electron captures at lower pressures and so the density drop would occur at higher altitudes in the hotter parts of the crust. This is known as a \textit{thermal mountain} \citep{1998ApJ...501L..89B, 2000MNRAS.319..902U, 2005ApJ...623.1044M, 2006MNRAS.373.1423H, 2006ApJ...641..471P, 2013PhRvD..88d4004J}. Another mechanism through which mass-quadrupoles can be built are through mountains sustained by magnetic stresses, called \textit{magnetic mountains} \citep{2002PhRvD..66h4025C, 2008MNRAS.385..531H}. These can occur when a NS has a sufficiently large toroidal or poloidal magnetic field which will act to distort the NS into an oblate or prolate shape and will naturally produce a quadrupole if the spin- and magnetic-axes are misaligned. A third way through which GWs can arise is through internal \textit{r}-mode instabilities \citep{1998ApJ...502..708A, 1999ApJ...516..307A, 1999ApJ...517..328L, 2000ApJ...534L..75A, 2002MNRAS.337.1224A, 2002ApJ...574L..57H, 2002ApJ...578L..63W, 2006PhRvD..73h4001N, 2007PhRvD..76f4019B}. In a perfect fluid, these modes are unstable for all rates of rotation due to GW emission.

We explored whether GWs could explain the observed distribution and, if so, whether there is a preference for any of the GW-production mechanisms. For our analysis, we did not consider mountains solely created by the magnetic field, nor did we consider magnetic mountains built through accretion. For these cases, the magnetic fields are not strong enough to sustain sufficiently large mountains for the spin evolution of these systems to be noticeably affected \citep[see][]{2008MNRAS.385..531H, 2011MNRAS.417.2696P, 2017PhRvL.119p1103H}.

\section{Spin-Evolution Model}
\label{sec:SpinModel}

For this work, we constructed a model for the spin evolution of an accreting NS. We incorporated the accretion-magnetosphere coupling by using the model of \citet[][; Equation~\ref{eq:HoPdot}]{2014MNRAS.437.3664H} and included a torque due to GW spin-down (Equation~\ref{eq:GWPdot}). The spin rate is a first-order time derivative and so  can be evolved numerically. Our assumed canonical values for a LMXB are shown in Table~\ref{tab:CanonicalLMXB}. For our canonical NS we did not include GW effects. We assumed constant density for our NSs, which affected the moment of inertia. For simplicity, we did not model the magnetic-field evolution. The time a NS is evolved for is denoted as the evolution time.

\begin{table*}
	\caption{Canonical values for a LMXB.}
	\label{tab:CanonicalLMXB}
	\begin{tabular}{c c c c c c c}
		\hline\hline
		Mass / \si{\solarMass} & Radius / \si{\kilo\metre} & $B$ / \si{\gauss} & Initial spin period / \si{\second} & $\xi$ & $\langle \dot{M} \rangle$ / \si{\solarMass\per\year} & $Q_{2 2}$ / \si{\gram\centi\metre\squared} \\
		\hline
		$1.4$ & $10$ & $10^8$ & $0.1$ & $0.5$ & $\num{5e-11}$ & $0$ \\
		\hline
	\end{tabular}
\end{table*}

Our model can evolve both persistent and transient accretors. For transient accretors we used a fast-rise, exponential-decay function, described in Section~\ref{sec:TransientAccretion} by Equations~(\ref{eq:AccretionProfile}--\ref{eq:AccretionRate}), and evolved the time-averaged spin-derivative, $\langle \dot{P}(P, \dot{M}) \rangle$, which, for a given NS, is a function of the spin and accretion rate. This average was obtained by averaging the spin derivative over one outburst/quiescence cycle. The time-average was evolved rather than the instantaneous spin rate, $\dot{P}(P, \dot{M})$, to simplify the integration procedure. Otherwise the integration procedure would have needed to take into account the full fast-rise, exponential-decay features of the accretion profile. For persistent accretors this was not a problem and so we could simply evolve $\dot{P}(P, \dot{M})$. Unless specified otherwise we used the following values for the transient accretion profile: $F_\text{t} = \SI{10}{\year}$, $T_\text{recurrence} = \SI{100}{\year}$ and $f_\text{max} / f_\text{min} = \num{e4}$. This was chosen for simplicity and to limit the explorable parameter space. Most of our simulations turned out to be relatively insensitive to the exact values of these parameters. Of course, should one be interested in modelling individual systems with this profile then particular care would need to be taken when tuning these parameters.

\begin{figure}
	\includegraphics[width=\columnwidth]{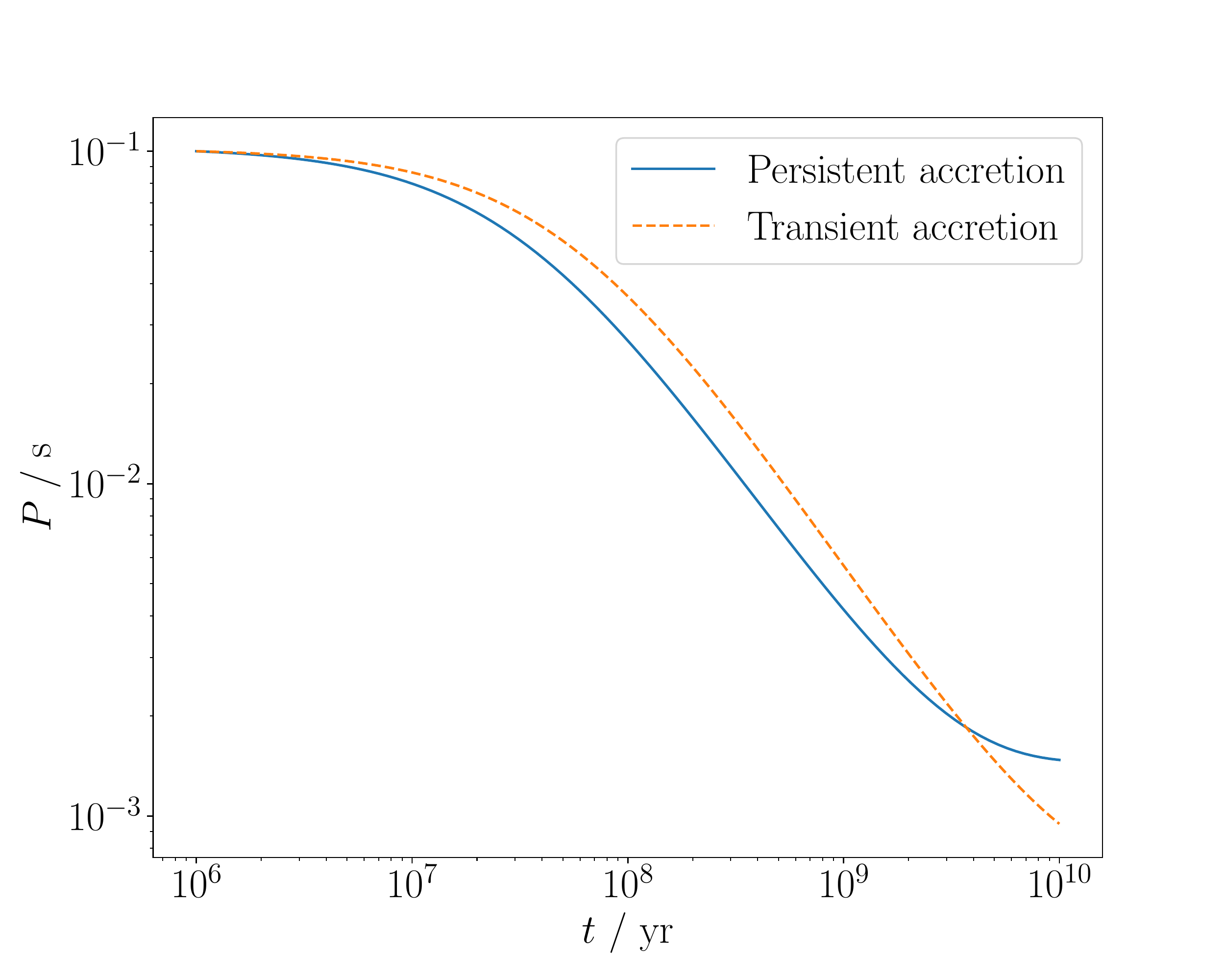}
	\caption{The spin evolution of the canonical accreting NS with persistent accretion (blue continuous line) and transient accretion (orange dashed line) with initial values from Table~\ref{tab:CanonicalLMXB}. The persistent accretor initially spins up faster than the transient accretor. However, towards the end of its evolution the persistent accretor begins to slow down and the transient accretor overtakes and reaches a faster final spin.}
	\label{fig:CanonicalEvolution}
\end{figure}

Figure~\ref{fig:CanonicalEvolution} shows the spin evolution of the canonical accreting NS with persistent and transient accretion. As was found by \citet{2017MNRAS.470.3316D} and \citet{2017ApJ...835....4B}, we see that the persistently-accreting NS initially spins up faster and reaches a final spin of $\nu = \SI{678}{\hertz}$. The transient system spins up slower but obtains a faster final spin of $\nu = \SI{1055}{\hertz}$. However, neither of the systems were evolved long enough to reach spin equilibrium. The upper limit of $\SI{e10}{\year}$ for the evolution time was chosen since no system can evolve for longer than the age of the Universe.

\section{Simulated Populations}
\label{sec:SimulatedPopulations}

In order to obtain a distribution of spins with which to compare to the observed distribution, we used a Monte Carlo population-synthesis method to draw the initial parameters from a given set of distributions and evolve each NS \citep[see][ for another NS population synthesis study]{1998ApJ...497L..97P}. Each NS was assigned a mass, radius, magnetic-field strength, initial spin period, average accretion rate, mass-quadrupole moment and evolution time. We evolved $1000$ NSs in each simulation.

\begin{table*}
	\caption{Initial values and evolution parameters for population synthesis.}
	\label{tab:InitialValues}
	\begin{tabular}{ l c l }
		\hline\hline
		\multicolumn{1}{c}{Parameter} & \multicolumn{1}{c}{Distribution} & \multicolumn{1}{c}{Values} \\ 
		\hline
		Mass / \si{\solarMass} & Single-value & $1.4$ \\ 
		Radius / \si{\kilo\metre} & Single-value & $10$ \\ 
		$\log_{10} (B \, / \, \si{\gauss})$	& Gaussian	& $\mu = 8.0$, $\sigma = 0.1$ \\
		Initial spin period / \si{\second}	& Flat	& $0.01-0.1$ \\
		$\xi$	& Single-value & $0.5$ \\
		$\log_{10}(\langle \dot{M} \rangle \, / \, \si{\solarMass\per\year})$ & Gaussian & $\mu = -11.0 + \log_{10}(5)$, $\sigma = 0.1$ \\
		$Q_{2 2}$ / \si{\gram\centi\metre\squared} & Single-value & $0$ \\
		Evolution time / \si{\year} & Flat-in-the-log & $10^9 - 10^{10}$ \\
		\hline
	\end{tabular}
\end{table*}

The first simulations were evolved using the distributions shown in Table~\ref{tab:InitialValues}. We fixed the masses and radii at $\SI{1.4}{\solarMass}$ and $\SI{10}{\kilo\metre}$, respectively, to match the canonical values for NSs. Typically, AMXPs are measured to have magnetic fields of $\sim \SI{e8}{\gauss}$ and so the field strength was taken from a log-Gaussian distribution with mean $\mu = 8.0$ and standard deviation $\sigma = 0.1$. The initial spin period was drawn from a flat distribution between $\SIrange{0.01}{0.1}{\second}$, which our simulations turned out to be relatively insensitive to. The correction factor $\xi$ was chosen to be $0.5$. The average accretion rate was motivated by observations of LMXBs and was given by a log-Gaussian with $\mu = -11.0 + \log_{10}(5)$ and $\sigma = 0.1$. For the initial simulations we assumed there was no GW component. We found that for evolution times much less than $\SI{e9}{\year}$ the NSs would not have enough time to spin up to frequencies above $\SI{100}{\hertz}$ and so the evolution time was taken from a flat-in-the-log distribution between $\SIrange{e9}{e10}{\year}$. The distribution was chosen to be flat-in-the-log in order for it to be scale-invariant \citep[as was used by][]{1998ApJ...497L..97P}, thus parametrising our uncertainty in the value of the evolution time.

\begin{figure*}
	\subfigure{\makebox[\columnwidth][c]{\includegraphics[width=1.1\columnwidth]{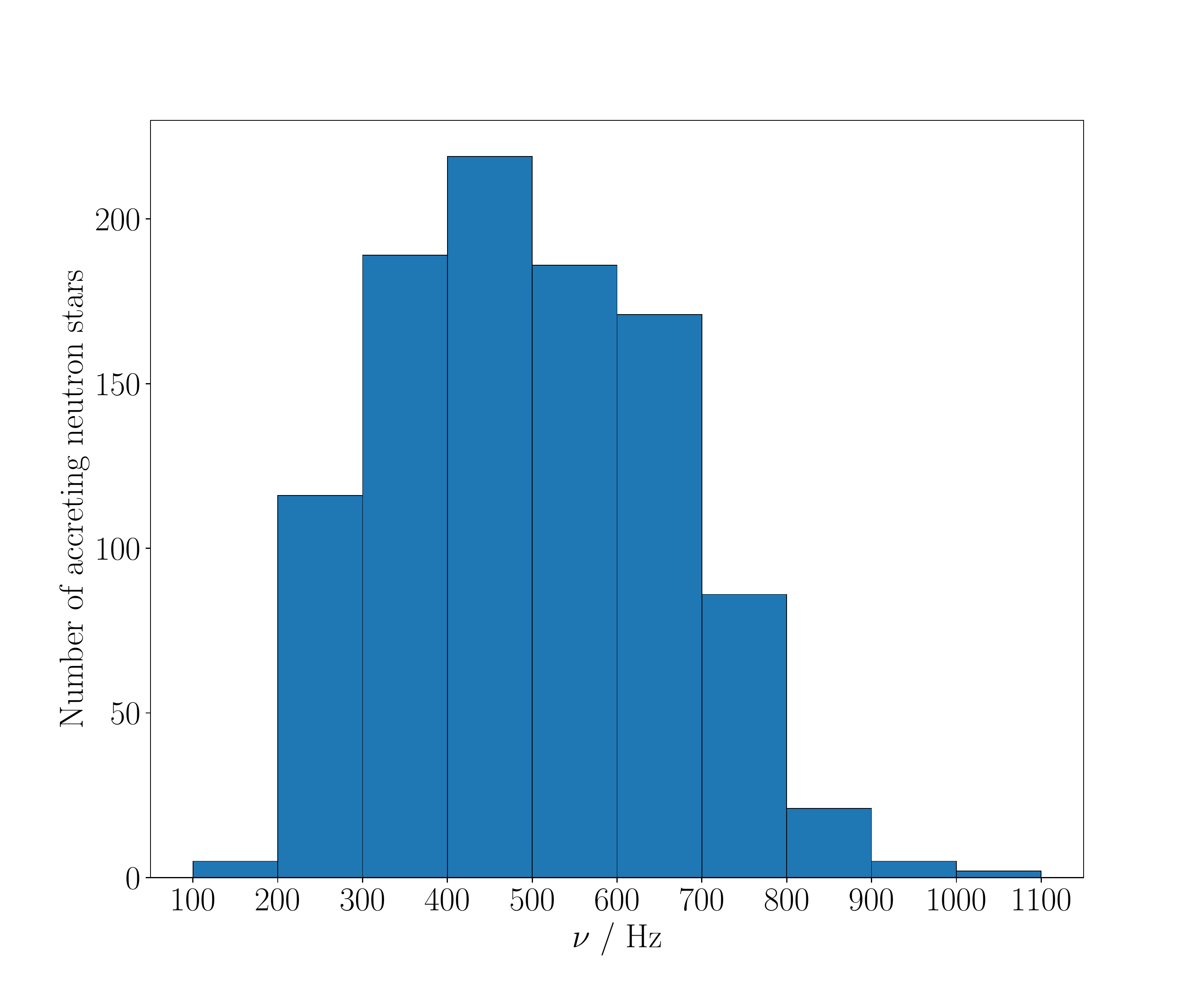}}}
	\subfigure{\makebox[\columnwidth][c]{\includegraphics[width=1.1\columnwidth]{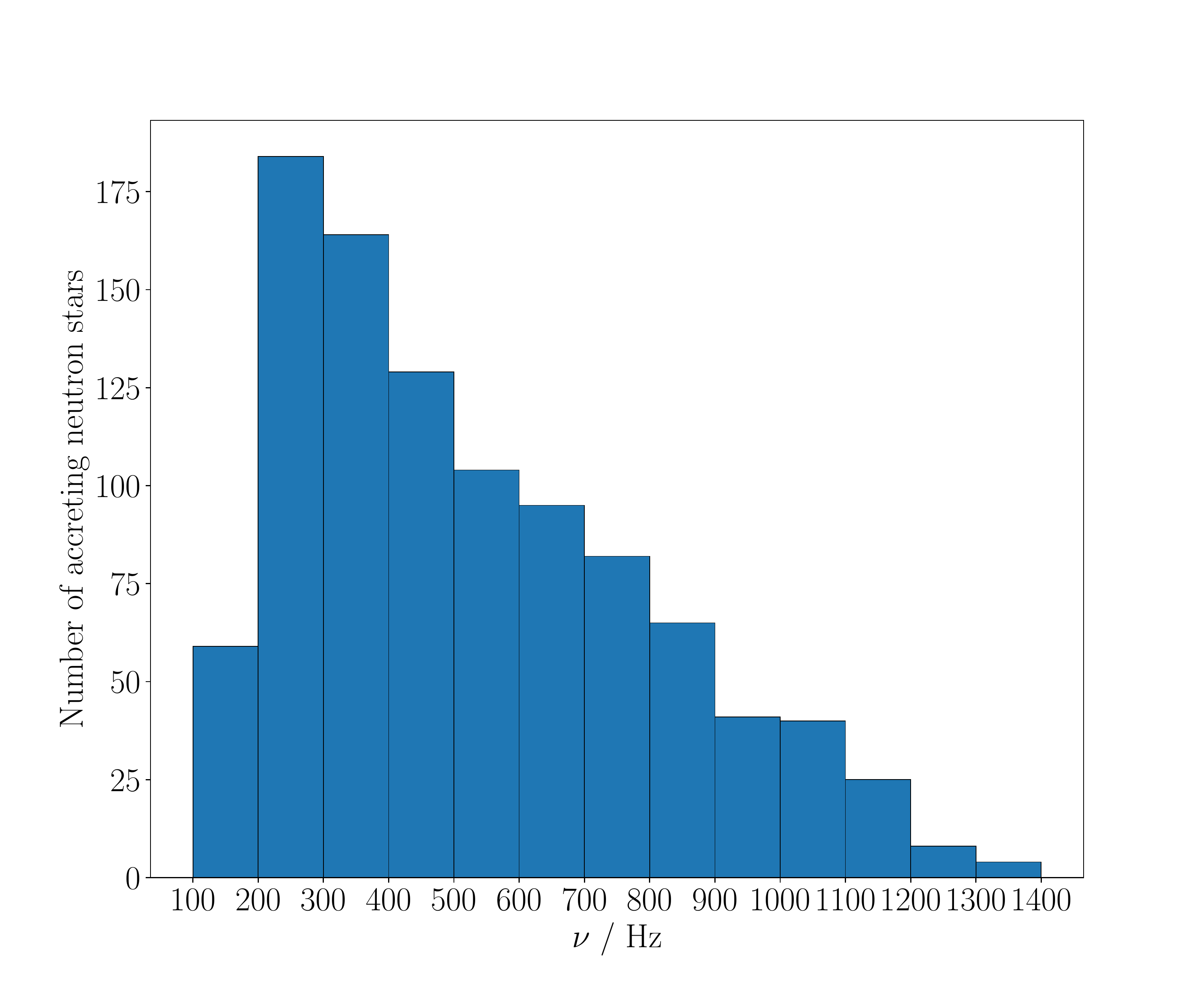}}}
	\caption{Distributions of spin frequencies for simulated persistently-accreting NSs (left panel) and transiently-accreting NSs (right panel) with initial distributions from Table~\ref{tab:InitialValues}.}
	\label{fig:NoQSpins}
\end{figure*}

The resultant spin-frequency distributions for persistent and transient accretors are shown in Figure~\ref{fig:NoQSpins}. One can see for this simple case that for both persistent and transient accretion we do indeed obtain NSs that spin in excess of $\SI{1}{\kilo\hertz}$. More generally, we observe that we get many NSs that spin faster than the observed spin-frequency limit of $\sim \SI{600}{\hertz}$ and, as one might expect, we find more high-frequency NSs for the transient case. This is because transient accretion enables these stars to spin to higher frequencies than with persistent accretion, provided they evolve for long enough. For both simulations we have not obtained the characteristic behaviour of the observed distribution since there is no evidence for a pile-up of NSs at high frequencies.

In order to quantify how different our simulated populations are to the observed population we applied a Kolmogorov-Smirnov test to the distributions. This test allows one to compare two distributions by testing the null hypothesis that the two distributions are the same. We chose a significance level of $\alpha = 0.10$ for this work, which meant that should we have found $p$-values less than this value we could reject the null hypothesis with 90\% certainty for that case. For the persistent accretors we obtained a $p$-value of $p = \num{7.7e-2}$ and for the transient accretors we obtained $p = \num{9.9e-4}$. This meant that we could reject the null hypothesis at the 10\% significance level that the observed distribution is drawn from the persistent population or the transient population. 

\begin{figure}
	\subfigure{\makebox[\columnwidth][c]{\includegraphics[width=1.1\columnwidth]{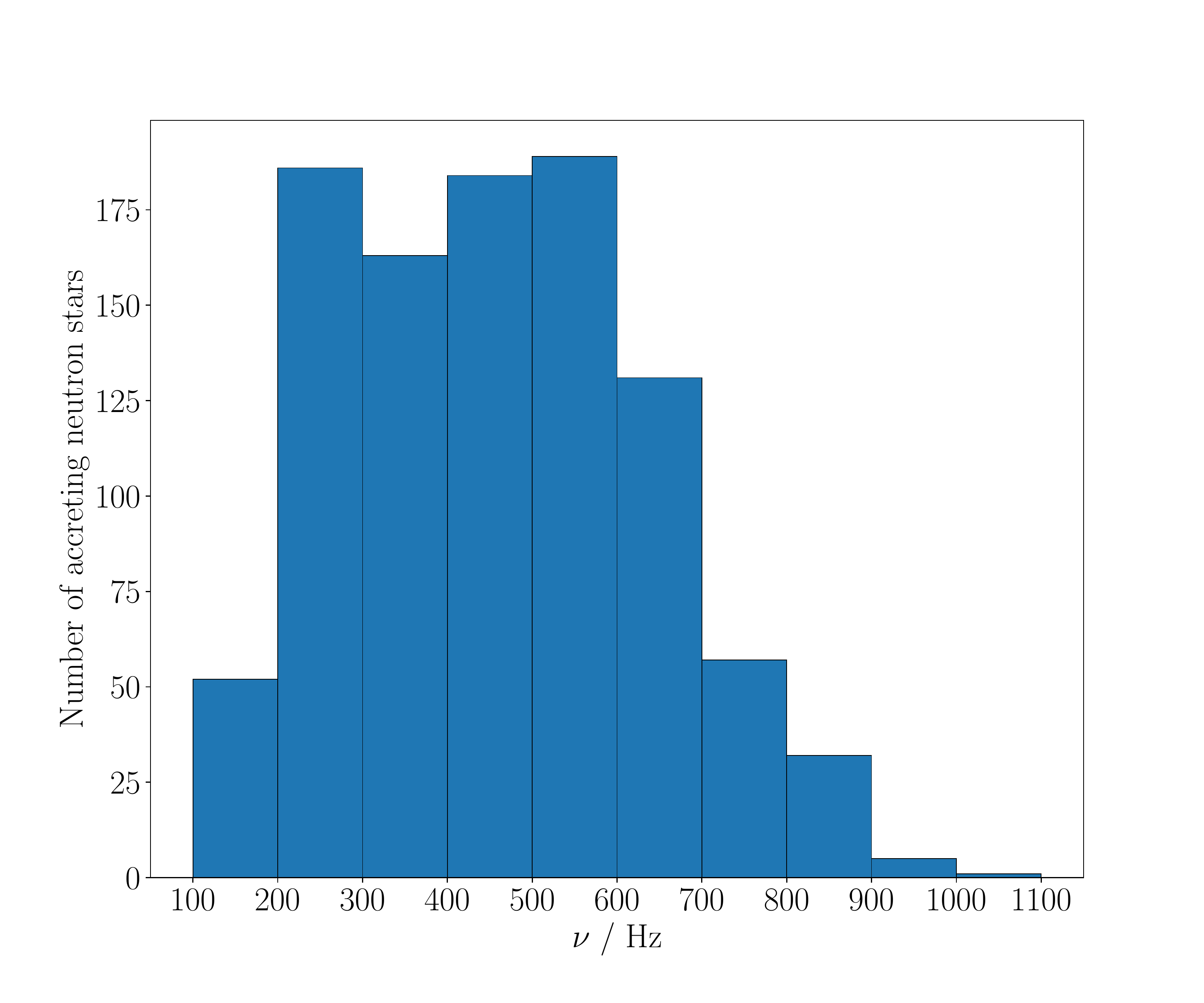}}}
	\caption{Distribution of spin frequencies for simulated transiently-accreting NSs with initial distributions from Table~\ref{tab:InitialValues} including the magnetic-dipole torque.}
	\label{fig:NoQSpinsEM}
\end{figure}

We explored the effect that magnetic-dipole radiation (Equation~\ref{eq:EMPdot}) has on transient accretors using the initial distributions in Table~\ref{tab:InitialValues}. The results are displayed in Figure~\ref{fig:NoQSpinsEM}. The inclusion of this additional torque stops many of the systems from spinning up to sub-millisecond periods. We obtain $p = \num{0.20}$ so we cannot rule out the null hypothesis with any statistical certainty. However, in regards to the shape of the distribution, we do not obtain a sharp peak at the observed spin-frequency limit. Instead, we find a broad peak in the range $\SIrange{200}{600}{\hertz}$.

This demonstrates that a simple model for accretion is not sufficient to explain the observations of accreting NSs and also suggests that the inclusion of magnetic-dipole torques do not resolve this tension either. Therefore, an additional component needs to be included into the model.

\subsection{Including Gravitational-Wave Torques}
\label{sec:IncludingGWs}

\begin{figure*}
	\subfigure{\makebox[\columnwidth][c]{\includegraphics[width=1.1\columnwidth]{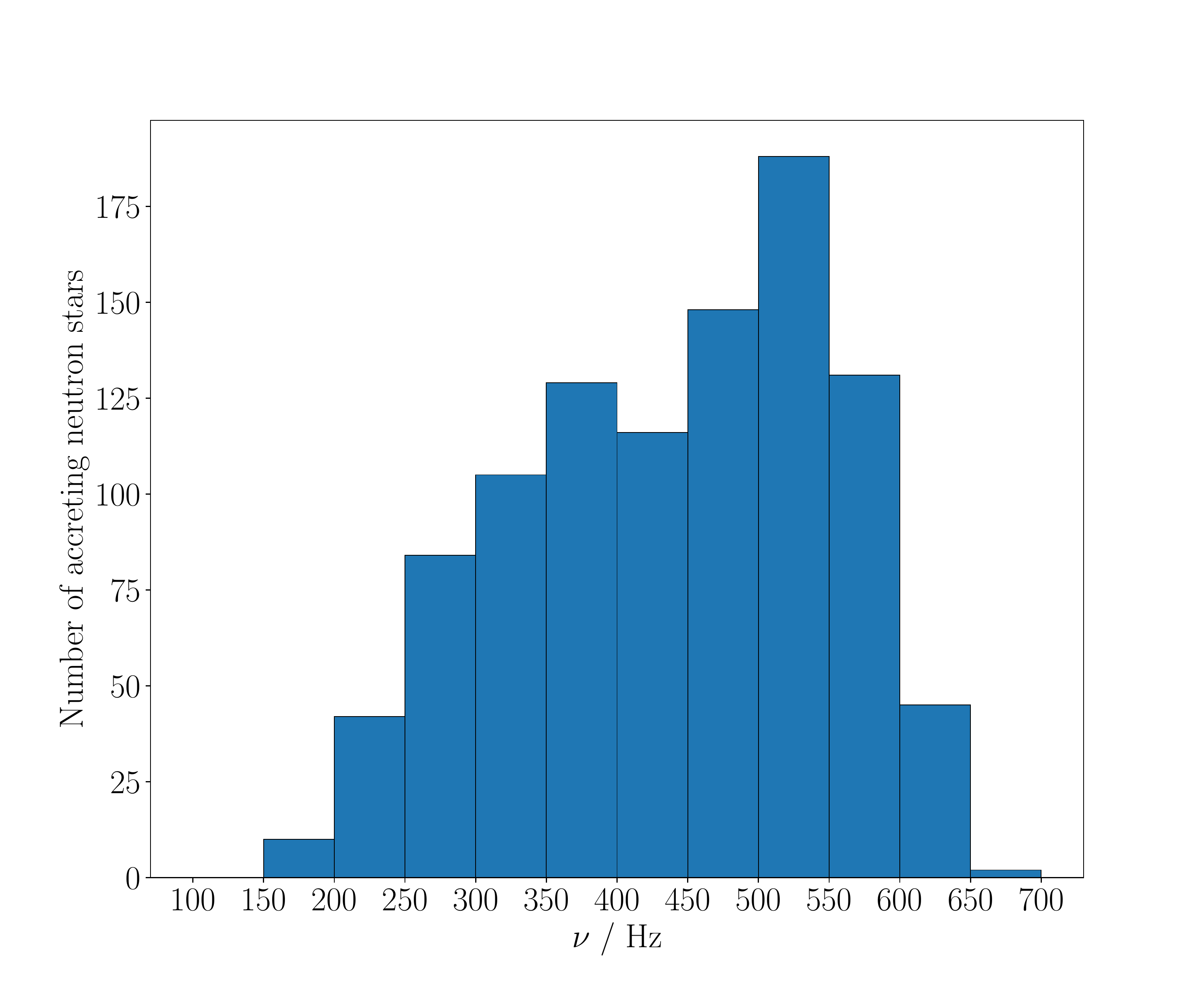}}}
	\subfigure{\makebox[\columnwidth][c]{\includegraphics[width=1.1\columnwidth]{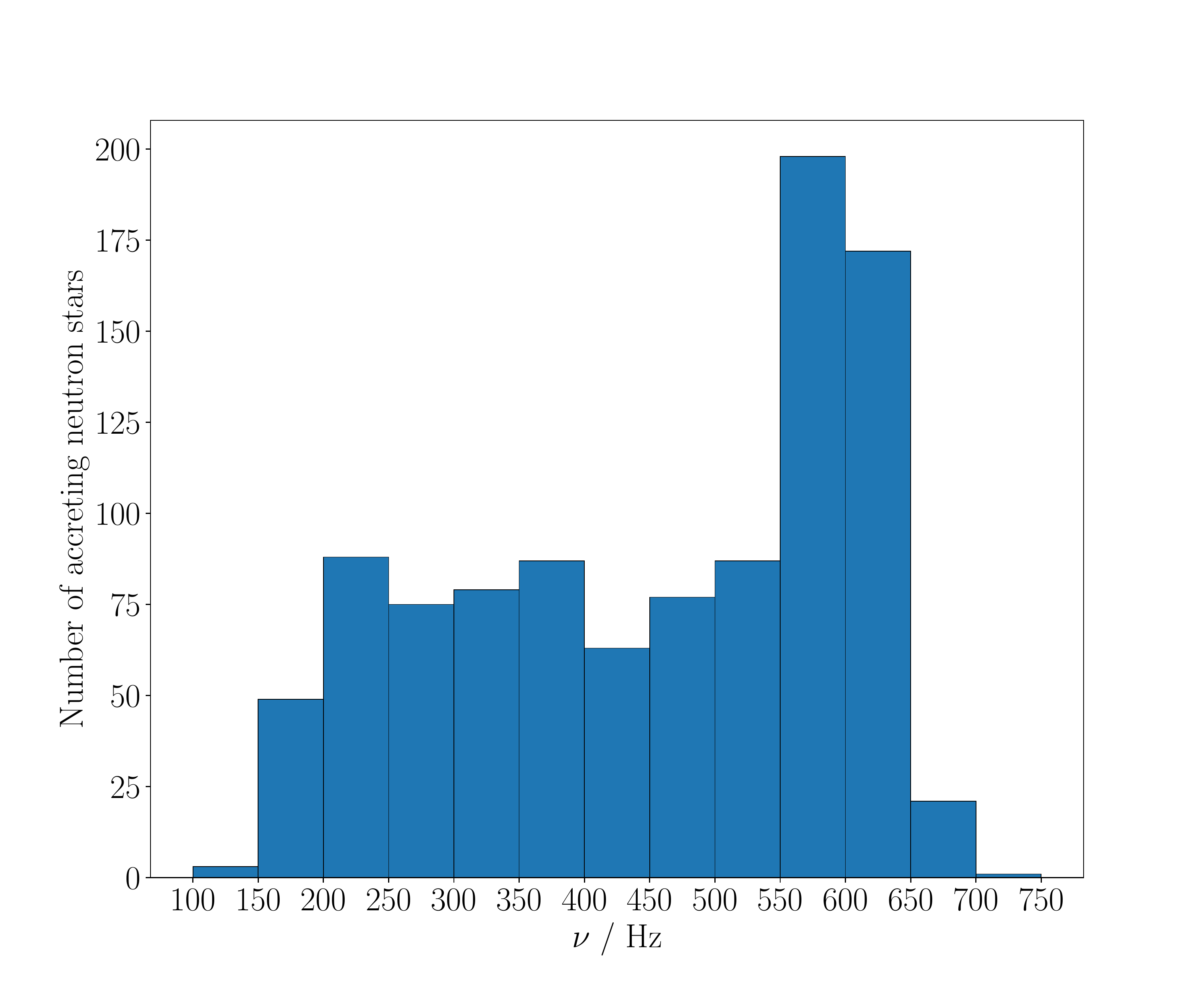}}}
	\caption{Distributions of spin frequencies for simulated persistently-accreting NSs (left panel) and transiently-accreting NSs (right panel) with initial distributions from Table~\ref{tab:InitialValues} and a fixed quadrupole of $Q_{2 2} = \SI{e36}{\gram\centi\metre\squared}$.}
	\label{fig:FixedQSpins}
\end{figure*}

We explored whether including a GW component to the spin evolution of accreting NSs could give us the observed spin distribution. Motivated by the necessary quadrupole in order to achieve torque balance (Equation~\ref{eq:BalanceQ}), we repeated the same simulations but with a fixed $Q_{2 2} = \SI{e36}{\gram\centi\metre\squared}$ for all NSs. (Physically, this can be interpreted as a permanent crustal mountain.) Figure~\ref{fig:FixedQSpins} shows the final-spin distributions for these simulations. This quadrupole has notably stopped the NSs from spinning up to sub-millisecond periods and has resulted in a pile-up centred on the $\SIrange{500}{550}{\hertz}$ bin for the persistent accretors and at $\SIrange{550}{600}{\hertz}$ for the transient accretors. This has appeared since the GW torque imposes a spin-frequency limit on the NSs. The peak for transient accretors is promising as this is where the peak lies for the spin distribution that we observe (cf. Figure~\ref{fig:SpinDistribution}). Interestingly, there is also a broader peak at lower frequencies. We found that $\approx 19\%$ of persistently-accreting NSs and $\approx 14\%$ of transiently-accreting NSs reached spin equilibrium by the end of the simulation. The systems that had reached spin equilibrium were clustered around the high-frequency peaks. We obtained $p = 0.19$ and $p = 0.80$ for the persistent and transient cases, respectively, and therefore were unable to reject the null hypothesis for both populations.

\begin{figure}
	\subfigure{\makebox[\columnwidth][c]{\includegraphics[width=1.1\columnwidth]{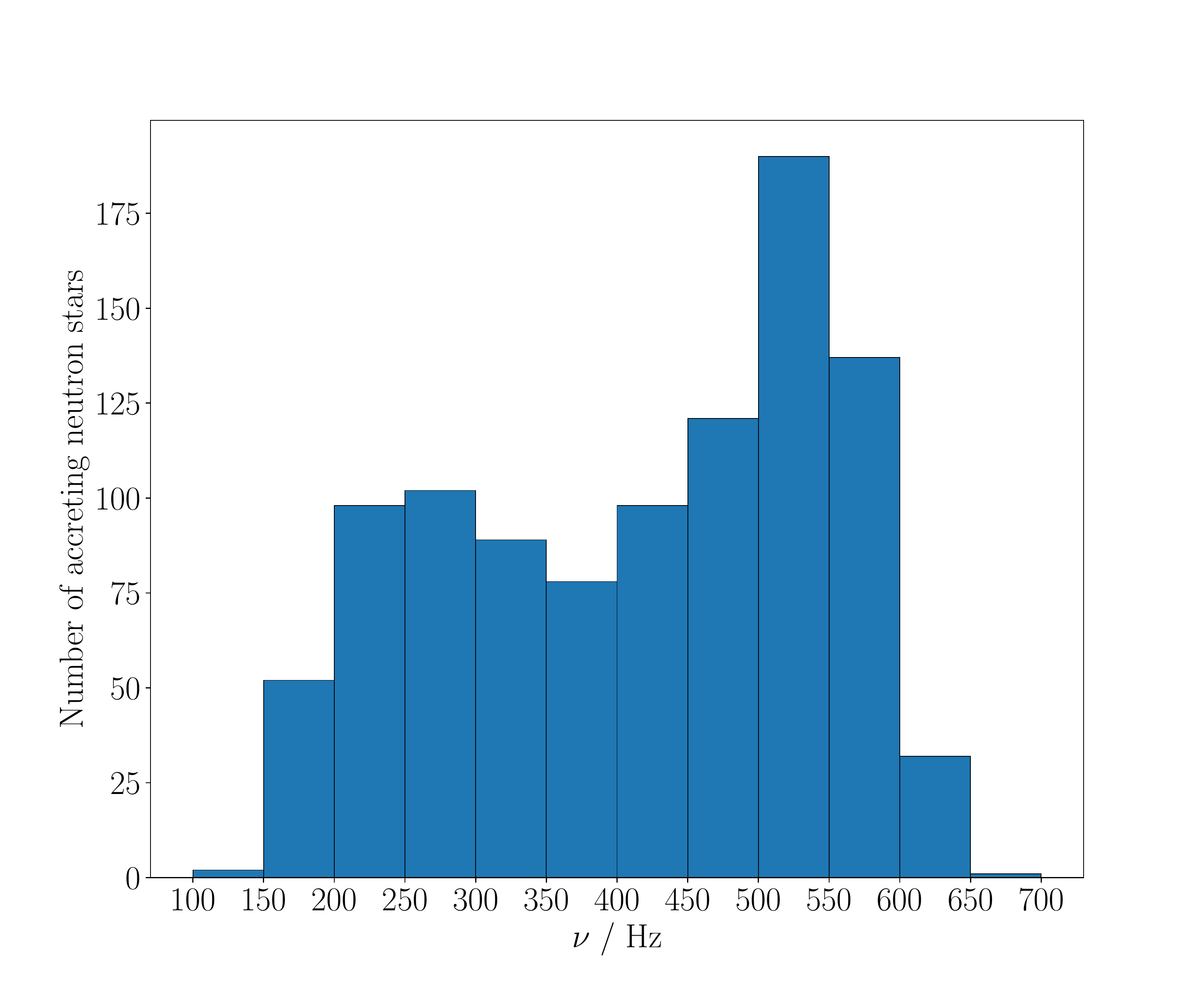}}}
	\caption{Distribution of spin frequencies for simulated transiently-accreting NSs with initial distributions from Table~\ref{tab:InitialValues} with a fixed quadrupole of $Q_{2 2} = \SI{e36}{\gram\centi\metre\squared}$ and including the magnetic-dipole torque.}
	\label{fig:QSpinsEM}
\end{figure}

We considered how magnetic-dipole radiation affects this picture for systems undergoing transient accretion. We used the same quadrupole and obtained the results shown in Figure~\ref{fig:QSpinsEM}. Interestingly, this distribution is qualitatively similar to the results without magnetic-dipole torques (right panel of Figure~\ref{fig:FixedQSpins}). We recover a pronounced peak at higher frequencies $\SIrange{500}{550}{\hertz}$, which by comparison has shifted down by $\SI{50}{\hertz}$. Since the features of the distribution remain the same we argue that one could obtain a distribution with a peak that matches the observed distribution through slight adjustment of the initial values, e.g., the quadrupole. Such an adjustment would be justifiable since there is significant uncertainty in many of these parameters. We could not reject the null hypothesis for these results with $p = 0.39$.

\begin{figure}
	\subfigure{\makebox[\columnwidth][c]{\includegraphics[width=1.1\columnwidth]{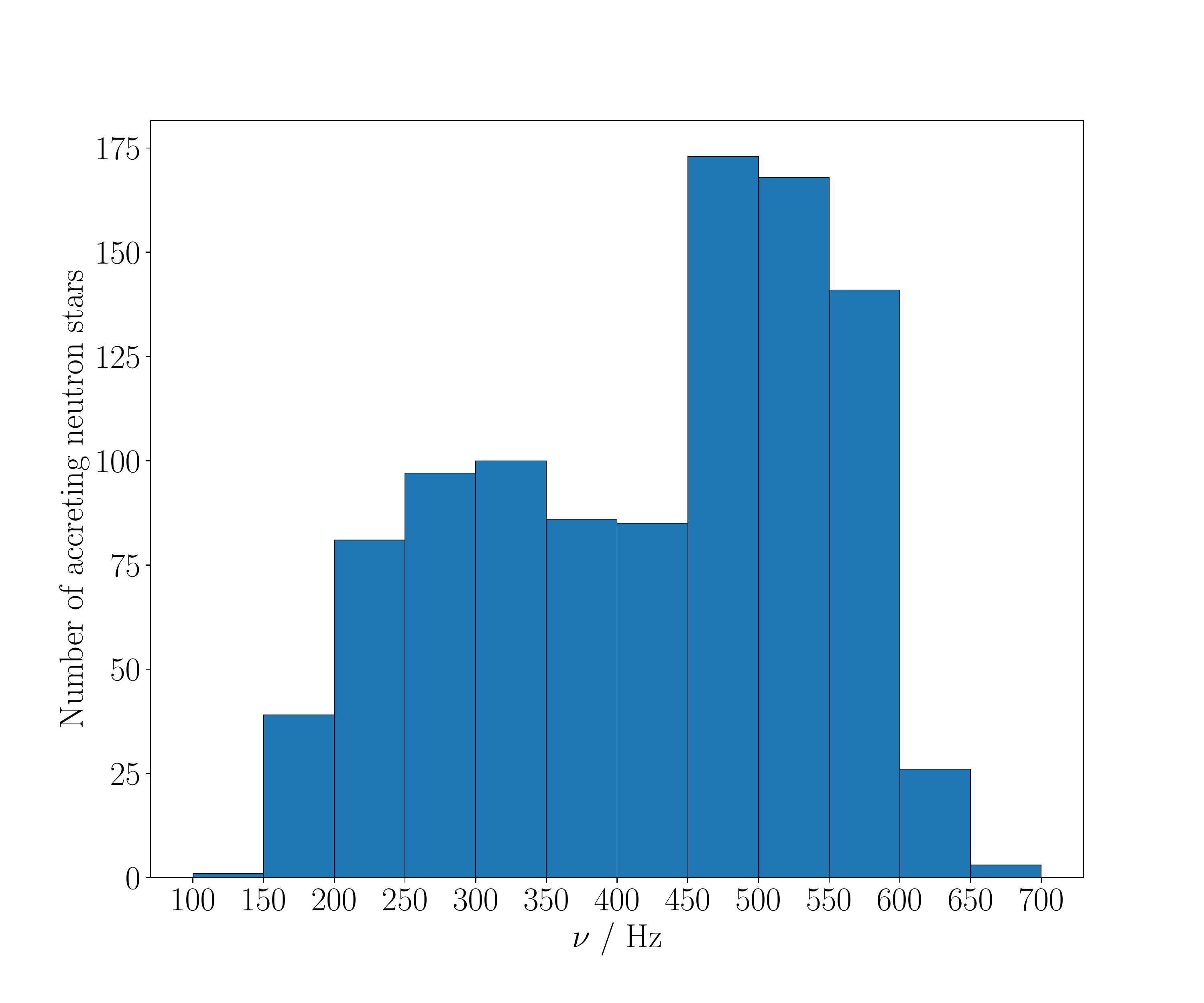}}}
	\caption{Distribution of spin frequencies for simulated transiently-accreting NSs with initial distributions from Table~\ref{tab:InitialValues} with a fixed quadrupole of $Q_{2 2} = \SI{e36}{\gram\centi\metre\squared}$, including the magnetic-dipole torque and $F_\text{f}$ distributed flat between $\SIrange{1}{100}{\year}$.}
	\label{fig:QSpinsEMF_t}
\end{figure}

For accreting NS systems the magnetic-dipole torque is expected to be negligible during outbursts, but it could play an important role during the quiescent phases. We explored a range of outburst durations, $F_\text{t}$ distributed uniformly between $\SIrange{1}{100}{\year}$, to assess the impact this made on the resultant spin distribution (Figure~\ref{fig:QSpinsEMF_t}). For this wide range of outburst lengths we find a broad peak between $\SIrange{450}{600}{\hertz}$. This contrasts the narrow peak in Figure~\ref{fig:SpinDistribution}. We obtain a \textit{p}-value of $p = 0.38$.

\begin{figure}
	\subfigure{\makebox[\columnwidth][c]{\includegraphics[width=1.1\columnwidth]{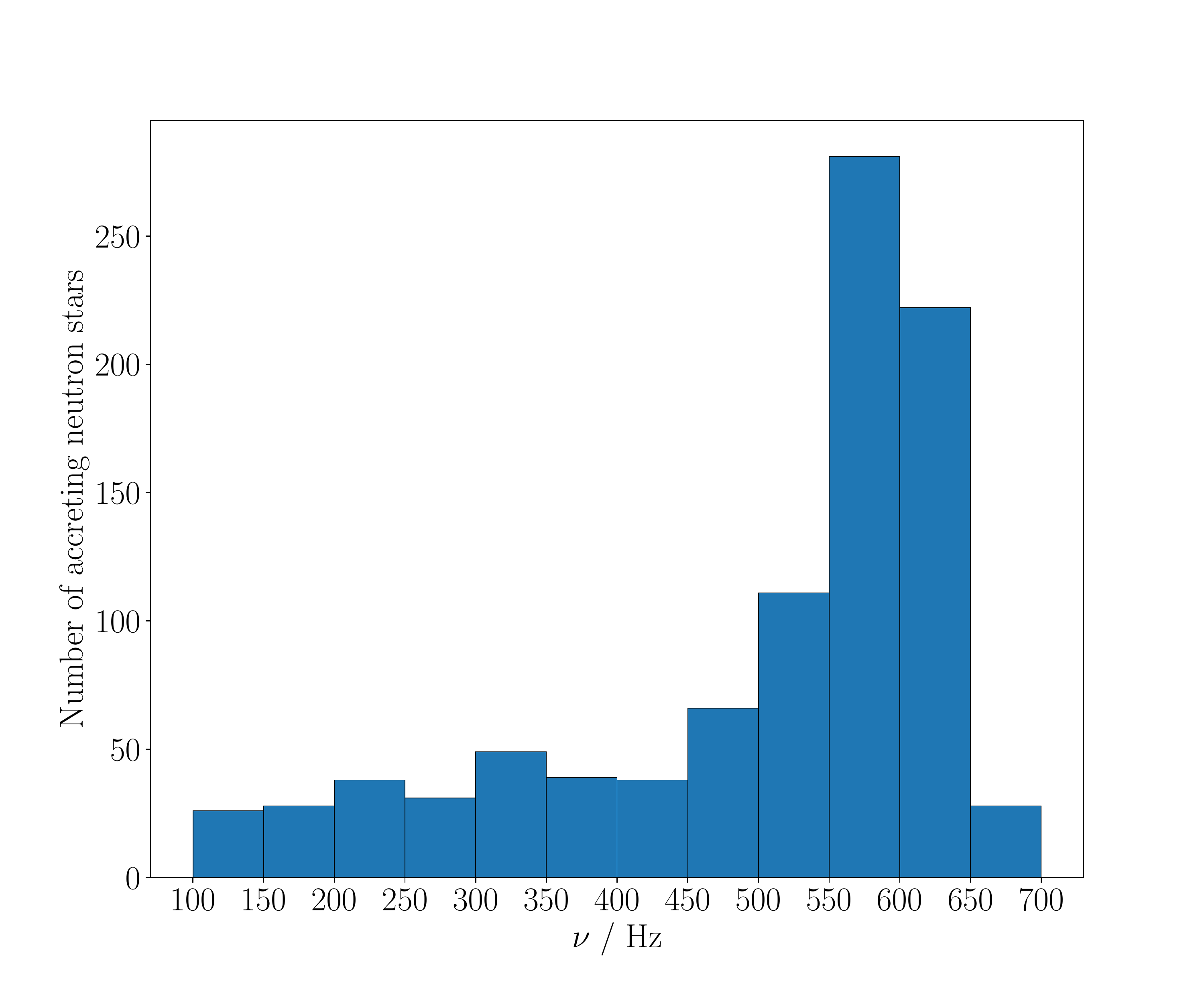}}}
	\caption{Distribution of spin frequencies for simulated transiently-accreting NSs with initial distributions from Table~\ref{tab:InitialValues} except for a fixed quadrupole of $Q_{2 2} = \SI{e36}{\gram\centi\metre\squared}$ and an evolution time distributed flat between $\SIrange{e8}{e10}{\year}$.}
	\label{fig:Ouch}
\end{figure}

We investigated how sensitive the simulated populations were on the distribution of the evolution time. We ran a simulation with the same quadrupole and the evolution time distributed flat between $\SIrange{e8}{e10}{\year}$ for transient accretors. This was motivated by assuming that NSs are born at a uniform rate, which is intuitively what one might expect, and that there are no selection effects to suggest that we are more likely to observe younger systems. The resultant spin-distribution is shown in Figure~\ref{fig:Ouch}. As was observed in the case when the time was distributed flat-in-the-log, there exists a pronounced peak towards the higher spin-frequencies. However, there are far fewer systems spinning at frequencies below this peak. For this simulation we obtained a $p$-value of $p = \num{1.2e-2}$ and thus could reject the null hypothesis. This distribution does not match what we observe.

\subsection{Thermal Mountains}
\label{sec:Thermal}

One of the most promising avenues for producing a mass-quadrupole on a fast-spinning, accreting NS is through thermal mountains built during accretion phases through asymmetries in pycnonuclear reaction rates \citep{2017PhRvL.119p1103H}. As a NS accretes matter composed of light elements, the matter becomes buried by accretion and is then compressed to higher densities. This causes the matter to undergo nuclear reactions such as electron captures, neutron emission and pycnonuclear reactions \citep{1990A&A...227..431H}. If the accretion flow is asymmetric this can cause asymmetries in density and heating which can produce a quadrupole moment. The quadrupole moment due to asymmetric crustal heating from nuclear reactions is approximated by \citep{2000MNRAS.319..902U}
\begin{equation}
	Q_{2 2} \approx \num{1.3e37} R_6^4 \left( \frac{\delta T_\text{q}}{\SI{e7}{\kelvin}} \right) \left( \frac{E_\text{th}}{\SI{30}{\mega\electronVolt}} \right)^3 \, \si{\gram\centi\metre\squared},
	\label{eq:ThermalQ}
\end{equation}
where $\delta T_\text{q}$ is the quadrupolar temperature increase due to the nuclear reactions and $E_\text{th}$ is the threshold energy for the reactions to occur. The value $\delta T_\text{q}$ will be a fraction of the total heating \citep{2001MNRAS.325.1157U},
\begin{equation}
\begin{split}
	\delta T &\approx \num{2e5} \left( \frac{C}{k_\text{B} \, \text{baryon}} \right)^{-1} \left( \frac{p}{\SI{e30}{\erg\per\centi\metre\cubed}} \right)^{-1} \\
	&\quad \times \left( \frac{Q}{\SI{1}{\mega\electronVolt}} \right) \left( \frac{\Delta M}{\SI{e-9}{\solarMass}} \right) \, \si{\kelvin},
\end{split}
	\label{eq:ThermaldeltaT}
\end{equation}
where $k_\text{B}$ is the Boltzmann constant, $C$ is the heat capacity, $p$ is the pressure, $Q$ is the heat released locally due the reactions and $\Delta M$ is the accreted mass. Some of this heating will be converted into the quadrupolar temperature increase, however, it is unclear quite how much will be converted. \citet{2000MNRAS.319..902U} estimate that $\delta T_\text{q} / \delta T \lesssim 0.1$, but, in reality, this ratio is poorly understood.

These thermal mountains are built during accretion outbursts. During quiescence phases, the deformations are washed away on a thermal timescale \citep{1998ApJ...504L..95B},
\begin{equation}
	\tau_\text{th} \approx 0.2 \left( \frac{p}{\SI{e30}{\erg\per\centi\metre\cubed}} \right)^{3 / 4} \, \si{\year}.
\end{equation}
If the system is in quiescence for longer than this timescale then the thermal mountain will be washed away and a new mountain will be built during the next outburst.

We implemented the expression for a quadrupole moment due to these reactions from Equation~(\ref{eq:ThermalQ}) and assumed the following values for our NSs \citep[as estimated by][ for the pulsar J1023+0038]{2017PhRvL.119p1103H}: $C / \text{baryon} \approx \num{e-6} k_\text{B}$, $E_\text{th} = \SI{30}{\mega\electronVolt}$, $p = \SI{e30}{\erg\per\centi\metre\cubed}$ and $Q = \SI{0.5}{\mega\electronVolt}$. This meant that the quadrupole due to these reactions was dependent only on the accreted mass $\Delta M$ and the fraction $\delta T_\text{q} / \delta T$. For this mechanism, we only considered transient accretion since in persistent accretion these mountains will not wash away, but, instead, will get progressively larger until the crust can no longer sustain them. This is effectively modelled through a fixed quadrupole that represents the largest mountain that can be built. 

\begin{figure*}
	\subfigure{\makebox[\columnwidth][c]{\includegraphics[width=1.1\columnwidth]{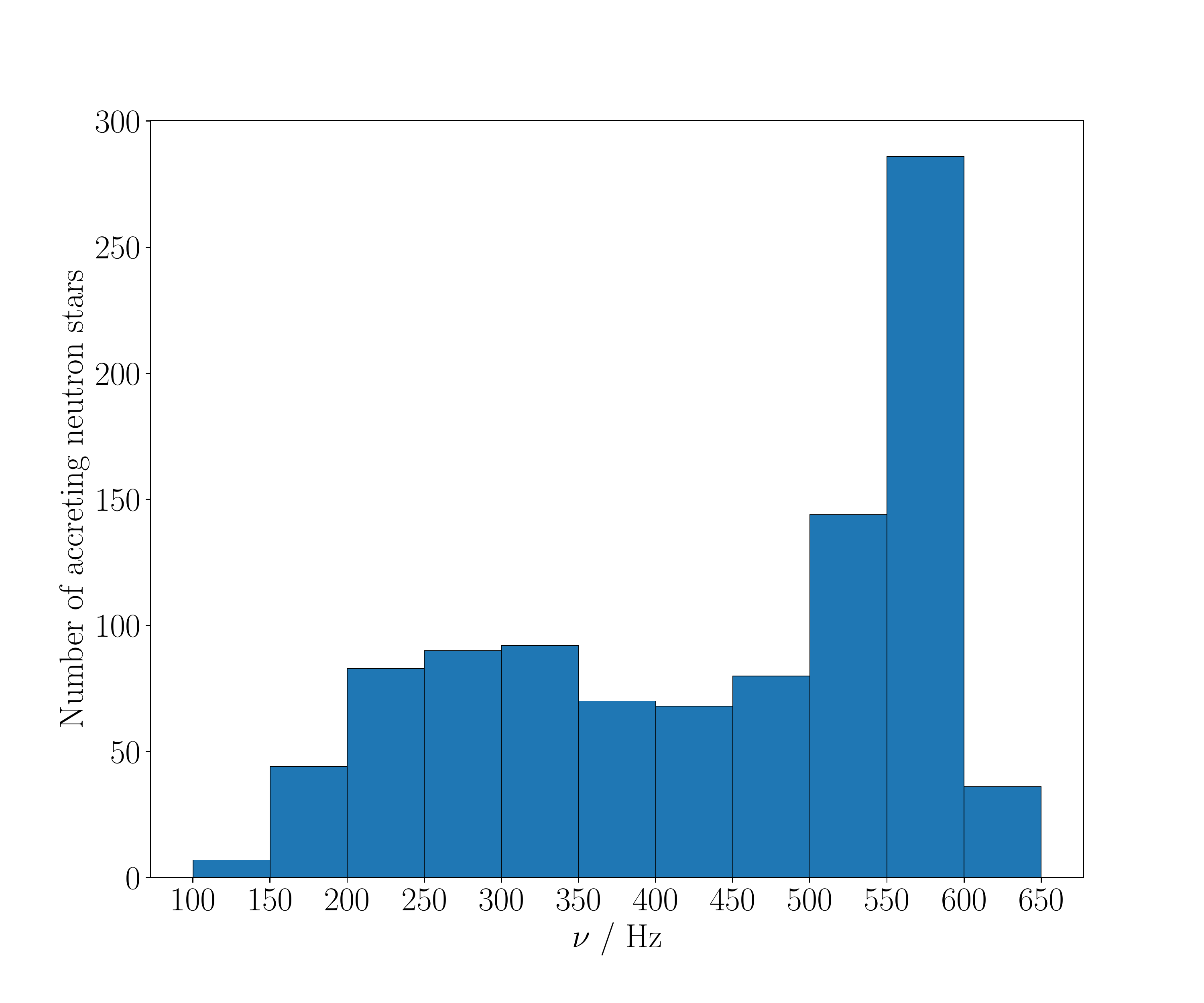}}}
	\subfigure{\makebox[\columnwidth][c]{\includegraphics[width=1.1\columnwidth]{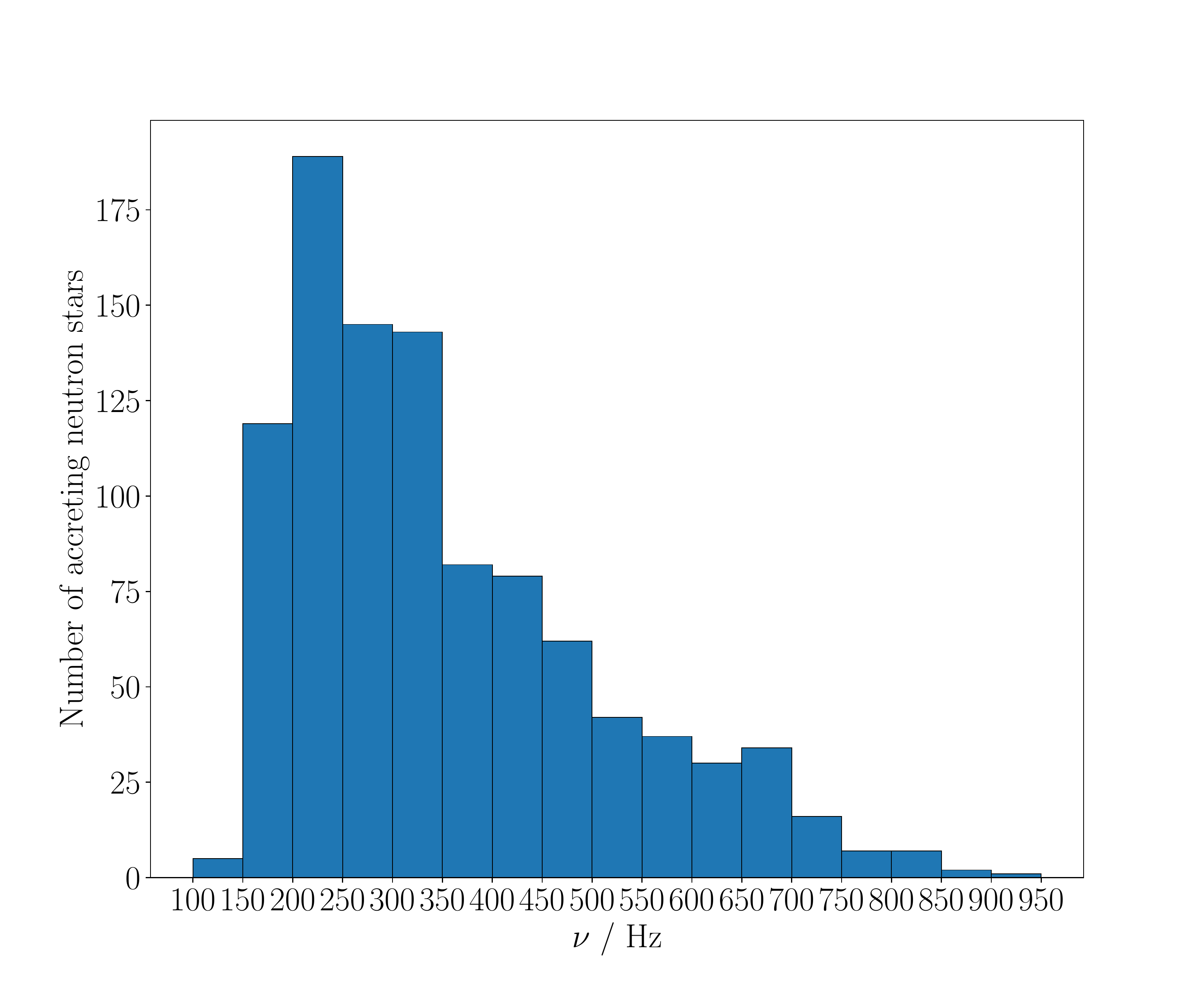}}}
	\caption{Distributions of spin frequencies for simulated transiently-accreting NSs that built thermal mountains during outburst phases with initial distributions from Table~\ref{tab:InitialValues}. The left panel has a fixed $\delta T_\text{q} / \delta T = \num{4e-4}$ and the right panel has $\delta T_\text{q} / \delta T$ distributed flat-in-the-log between $10^{-4}-10^{-2}$.}
	\label{fig:ThermalSpins}
\end{figure*}

In our model, we calculated $\Delta M$ by numerically integrating the accretion profile from the beginning of the outburst up to $F_\text{t}$. Our quiescence phases were long enough for the mountain to wash away during them. Unlike for our other prescriptions, we found that for thermal mountains, the specific values which parametrise the outburst features were very important in dictating the final-spin distribution. Based on observations of X-ray transients, we chose $F_\text{t} = \SI{0.1}{\year}$, $T_\text{recurrence} = \SI{2.0}{\year}$ and $f_\text{max} / f_\text{min} = \num{e4}$. We simulated NSs that built thermal mountains with $\delta T_\text{q} / \delta T = \num{4e-4}$. The left panel of Figure~\ref{fig:ThermalSpins} shows the resultant distribution of final spins. We can see qualitatively that this distribution has similar features to the fixed quadrupole case (see the right panel of Figure~\ref{fig:FixedQSpins}). Another promising aspect of the final-spin distribution is the prominence of the high-frequency peak. Like what is observed, this peak is narrow and much larger than the values in other frequency bins. We found a $p$-value of $p = 0.60$ for this distribution.

The only constraint on the ratio of quadrupolar to total heating comes from the non-detection of X-ray emission of quadrupolar flux perturbations during quiescence phases in LMXBs, which gives $\delta T_\text{q} / \delta T \lesssim 0.1$ \citep{2000MNRAS.319..902U}. Currently, there is no reason to believe that this fraction should be constant for all NSs. To account for our uncertainty in this fraction, we distributed $\delta T_\text{q} / \delta T$ flat-in-the-log between $10^{-4}-10^{-2}$. The result of this simulation is shown in the right panel of Figure~\ref{fig:ThermalSpins}. The distribution peaks at low frequencies and then falls off towards higher frequencies. We obtained $p = \num{7.0e-2}$ for this case which meant that we could reject the null hypothesis. This shows how this prescription favours $\delta T_\text{q} / \delta T$ being a fixed value. Seeking a physical explanation for this preference of $\delta T_\text{q} / \delta T$ being a fixed value as opposed to being distributed is beyond the scope of this paper and has been left for future work.

\subsection{Unstable \textit{r}-modes}
\label{sec:rModes}

An \textit{r}-mode is a fluid mode of oscillation for which the restoring force is the Coriolis force. \citet{1998ApJ...502..708A} demonstrated that gravitational radiation destabilises the \textit{r}-modes of rotating stars. These modes are generically unstable to GW emission \citep{1998ApJ...502..714F} and satisfy the Chandrasekhar-Friedman-Schutz instability, which facilitates the star finding lower energy and angular momentum configurations that allow the mode amplitude to grow \citep{1970PhRvL..24..611C, 1978ApJ...222..281F}.

The \textit{r}-mode instability has long been considered a potential mechanism for imposing a spin limit on NSs in LMXBs \citep{1999ApJ...516..307A}. The typical picture involves a NS being spun up through accretion until it enters the \textit{r}-mode instability window. This instability region depends primarily on the spin of the NS and its core temperature. Once a NS has entered this region, it will emit gravitational radiation and begin to spin down until it reaches stability. This is expected to occur on a timescale much shorter than the age of the system and should result in most LMXBs being stable. However, theoretical models for the \textit{r}-mode instability demonstrate that many of the observed accreting NSs, in fact, lie inside the instability window \citep{2011PhRvL.107j1101H}. This result would be consistent if the saturation amplitude for these systems was small, $\alpha \approx 10^{-8} - 10^{-7}$, but this is at odds with predictions which suggest that the amplitude should be several orders of magnitude higher than this \citep{2007PhRvD..76f4019B}.

\citet{1998PhRvD..58h4020O} described a phenomenological model for the evolution of \textit{r}-modes and the spin of the star. In this model the quadrupole moment for a constant-density NS that is unstable due to \textit{r}-modes is given by
\begin{equation}
	Q_{2 2} \approx \num{1.67e33} \left( \frac{\alpha}{\num{e-7}} \right) M_{1.4} R_6^3 \left(\frac{P}{\SI{1}{\second}}\right)^{-1} \, \si{\gram\centi\metre\squared}.
	\label{eq:RModeQ}
\end{equation}
An interesting feature of this expression is its dependence on the spin of the NS. As a NS spins faster the quadrupole moment grows. This is different to what is expected from mountains. In fact, \textit{r}-modes and mountains could be differentiated from one another through the scaling of the quadrupoles as well as the frequency of the emitted GWs; for mountains the GW frequency is $2 \nu$, whereas for \textit{r}-modes the frequency is $4 \nu / 3$.

\begin{figure*}
	\subfigure{\makebox[\columnwidth][c]{\includegraphics[width=1.1\columnwidth]{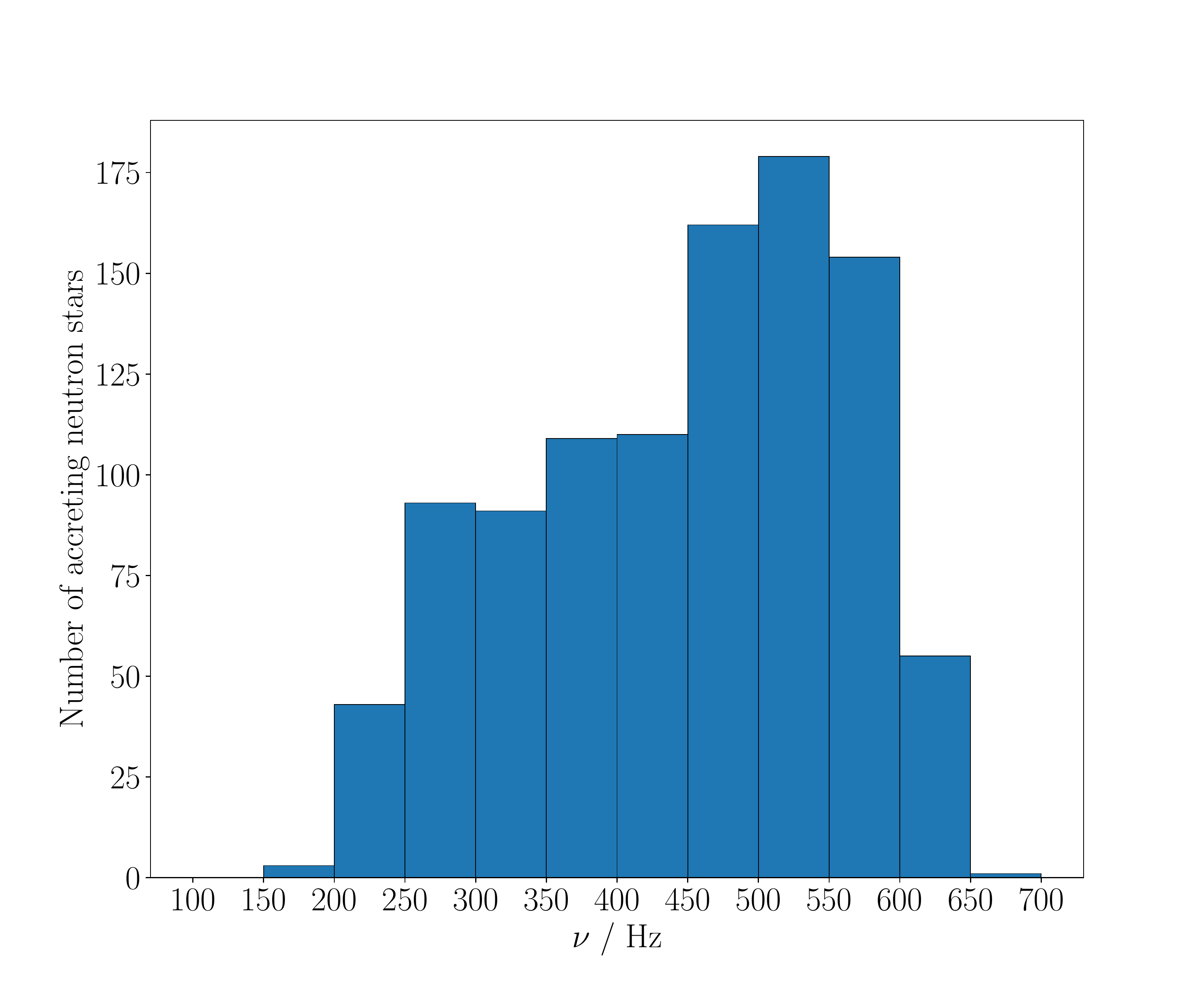}}}
	\subfigure{\makebox[\columnwidth][c]{\includegraphics[width=1.1\columnwidth]{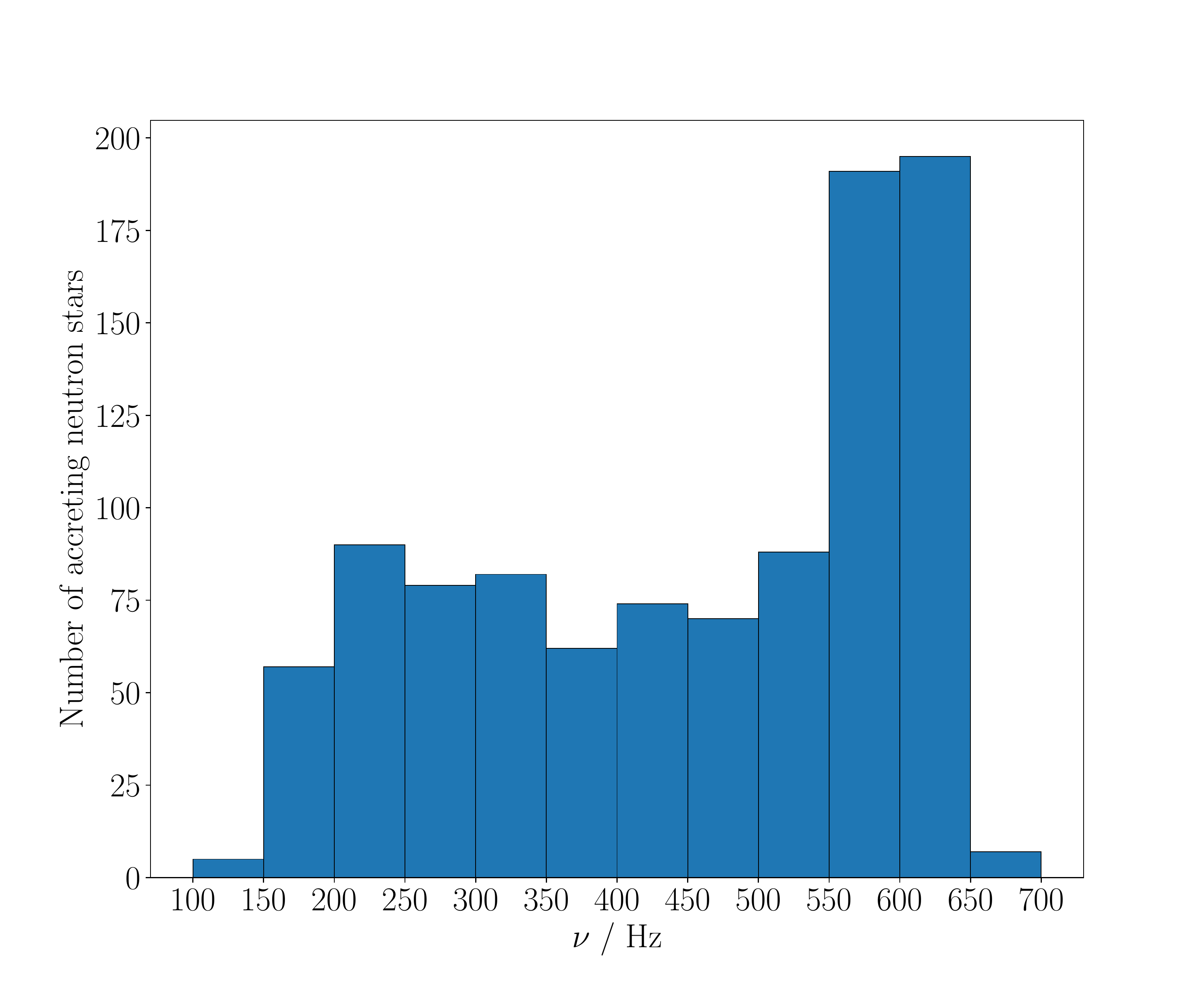}}}
	\caption{Distributions of spin frequencies for simulated persistently-accreting NSs (left panel) and transiently-accreting NSs (right panel) with unstable \textit{r}-modes with initial distributions from Table~\ref{tab:InitialValues} and $\alpha = \num{e-7}$.}
	\label{fig:RModeSpins1}
\end{figure*}

In order to simulate accreting NSs with unstable \textit{r}-modes, we implemented Equation~(\ref{eq:RModeQ}) into our model. We assumed that the mode-amplitude $\alpha$ remained constant for each NS. We repeated the previous simulations for persistent and transient accretors with unstable \textit{r}-modes and $\alpha = \num{e-7}$. Figure~\ref{fig:RModeSpins1} shows the final-spin distributions for those simulations. The unstable \textit{r}-modes were sufficient in both cases to give a peak at high spin-frequencies. For the persistently-accreting NSs, the peak was in the $\SIrange{500}{550}{\hertz}$ frequency bin, and for the transient accretors, the peak was in the $\SIrange{550}{600}{\hertz}$ bin. These distributions are similar to the case of a permanent quadrupole $Q_{2 2} = \SI{e36}{\gram\centi\metre\squared}$ (see Figure~\ref{fig:FixedQSpins}). For transient accretion with unstable \textit{r}-modes the peak is narrower and more pronounced indicating that the magnitude of the GW torque sets in quickly. This is due to the scaling of the quadrupole in Equation~(\ref{eq:RModeQ}), since it depends linearly on the spin. For the persistent case we found a $p$-value of $p = 0.28$ and for the transient case $p = 0.57$.

\begin{figure*}
	\subfigure{\makebox[\columnwidth][c]{\includegraphics[width=1.1\columnwidth]{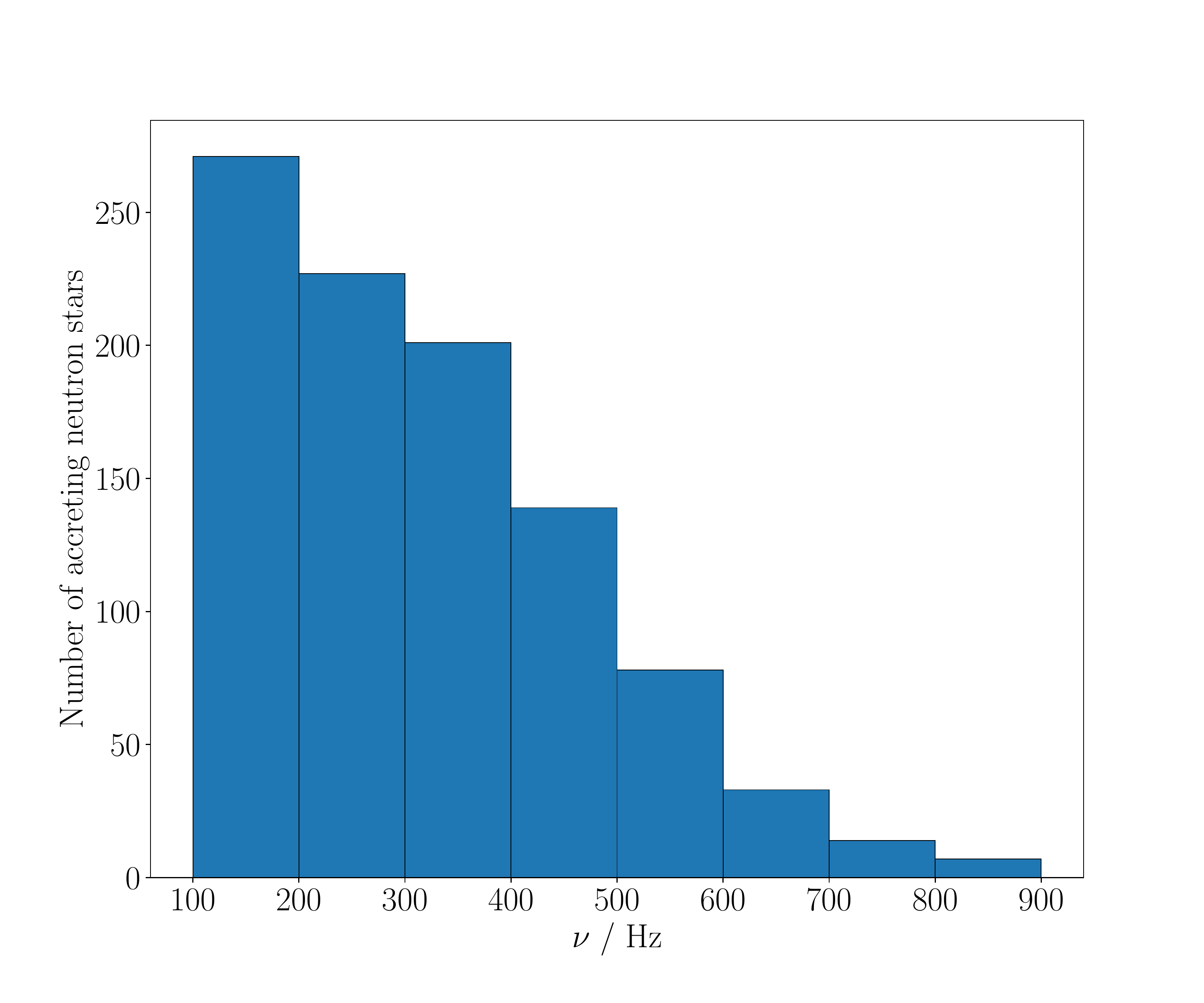}}}
	\subfigure{\makebox[\columnwidth][c]{\includegraphics[width=1.1\columnwidth]{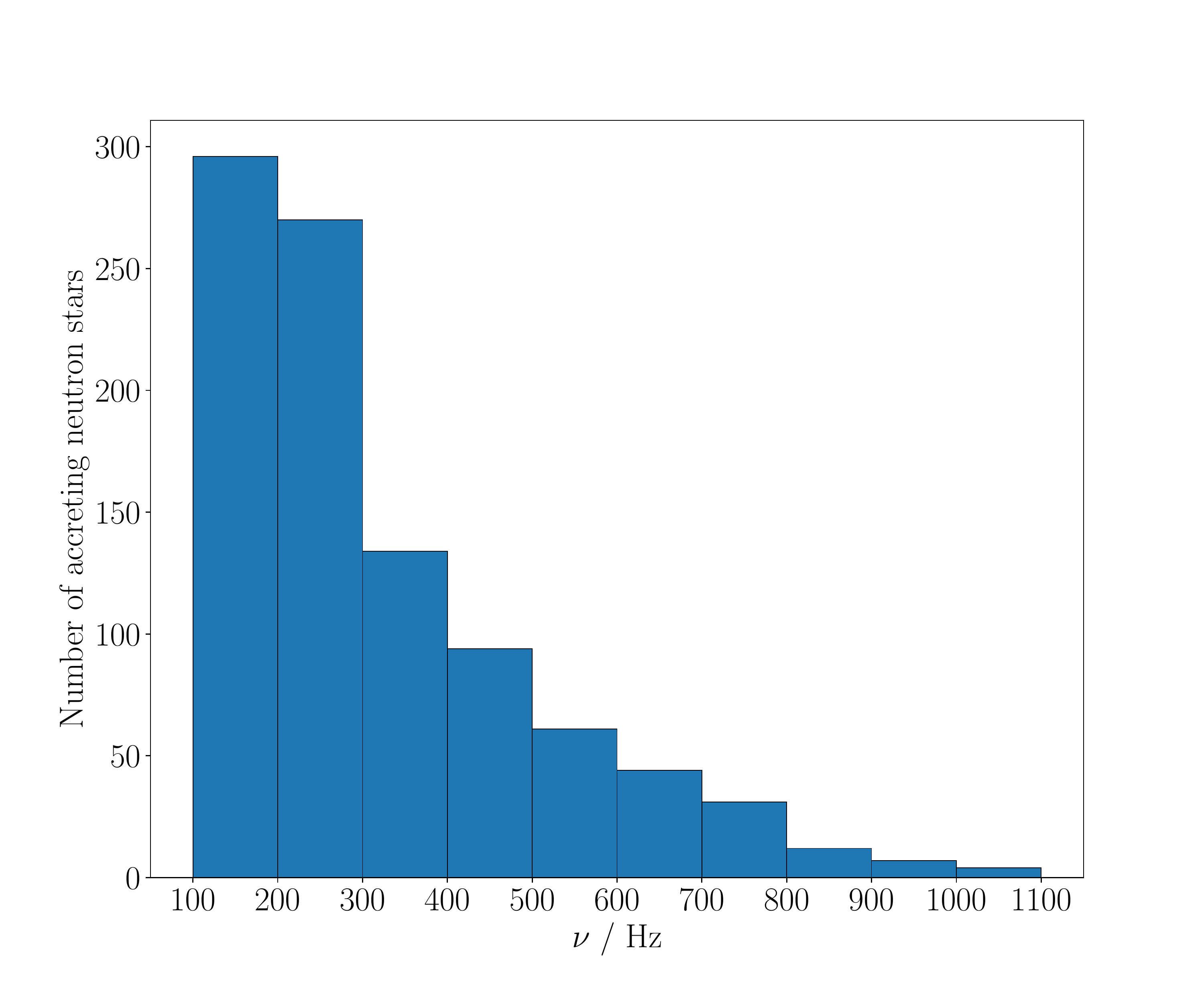}}}
	\caption{Distributions of spin frequencies for simulated persistently-accreting NSs (left panel) and transiently-accreting NSs (right panel) with unstable \textit{r}-modes with initial distributions from Table~\ref{tab:InitialValues} and $\alpha$ distributed flat-in-the-log between $10^{-8} - 10^{-4}$.}
	\label{fig:RModeSpins2}
\end{figure*}

We also conducted a simulation where $\alpha$ was distributed flat-in-the-log between $10^{-8} - 10^{-4}$. The result is shown in Figure~\ref{fig:RModeSpins2}. We can see, similar to the thermal mountain distribution, that both distributions follow an exponentially-decreasing behaviour. From these distributions we found $p = \num{1.2e-2}$ when the NSs were persistently accreting and $p = \num{1.5e-3}$ when they were transiently accreting. From these $p$-values we can reject the null hypothesis and note that the unstable-\textit{r}-modes prescription produces more promising results when $\alpha$ is fixed, which is in agreement with current theoretical expectations \citep{2003ApJ...591.1129A, 2007PhRvD..76f4019B}.

\section{Conclusions}
\label{sec:Conc}

An unresolved problem in the study of LMXBs is the unusual spin distribution of rapidly-accreting NSs and, in particular, why no NS has been observed to spin close to the centrifugal break-up frequency. A potential explanation to this problem comes from GWs. Theoretically, GWs could be able to spin down these systems away from the break-up frequency. However, there are a number of different mechanisms that could give rise to gravitational radiation and it is unclear which are the most probable. It is also unclear whether GWs are the only way to explain the observed distribution of accreting NSs. For example, it has recently been suggested by \citet{2016ApJ...822...33P} that spin-down torques from an enhanced pulsar wind due to a disc-induced opening of the magnetic field could have a meaningful effect on the spin evolution of an accreting neutron star. Such a torque is not phenomenologically accounted for in our method and could be a direction for future work.

In this paper, we have explored, within the context of our current understanding of accretion torques, whether an additional component is required in order to describe the spin evolution of accreting NSs. We investigated whether GW emission could be one such explanation and have compared competing GW mechanisms. We presented our model for the spin evolution of an accreting NS which accounts for accretion and magnetic-field effects, and also includes a GW spin-down component. Our model is able to simulate persistent and transient accretors.

In our simulations with no GW torques we obtained NSs with much higher spins than what is observed. We did not obtain any of the characteristic behaviour of the observed spin distribution. In particular, there was no evidence of a pile-up at high frequencies. However, by adding a permanent quadrupole moment, motivated by torque balance, of $Q_{2 2} = \SI{e36}{\gram\centi\metre\squared}$ we obtained qualitatively similar behaviour to the observed distribution for the transiently-accreting NS population.

We considered the impact of magnetic-dipole radiation on the above results. We found that in the case of no GW emission one does not obtain the observed distribution. With the inclusion of GW emission the resultant distribution is qualitatively similar to the case of no magnetic-dipole radiation. By varying the outburst duration with GW and magnetic-dipole torques we obtained a distribution with a broad high-frequency peak.

We investigated two GW-production prescriptions. For thermal mountains produced by asymmetric nuclear reactions in the crust, our model was sensitive to the precise features of the outburst profile, as well as the ratio of quadrupolar to total heating, $\delta T_\text{q} / \delta T$. We found that a value of $\delta T_\text{q} / \delta T = \num{4e-4}$ produced a similar distribution to what is observed. Promisingly, this gave the characteristic pile-up at high frequencies with a narrow, pronounced peak. We examined whether the distributions had a preference for $\delta T_\text{q} / \delta T$ being a single value or being distributed and found strong evidence arguing that it should be a fixed value. Accreting NSs with unstable \textit{r}-modes and $\alpha = \num{e-7}$ produced similar results to the case with a fixed quadrupole moment and the thermal mountain prescription. This prescription favoured $\alpha$ being fixed as opposed to being distributed.

The three cases that produced the most promising distributions that were qualitatively similar to the observed spin distribution -- permanent quadrupole, thermal mountains and unstable \textit{r}-modes -- are almost indistinguishable from one another. Although, the \textit{r}-mode-instability case could, in theory, be differentiated from the other prescriptions. This distinction could come from the fact that the quadrupole moment due to unstable \textit{r}-modes scales linearly with the spin frequency of the star. Another key difference comes from the frequency of the GWs that are emitted through this channel. Unstable \textit{r}-modes emit GWs with a frequency of $4 \nu / 3$, whereas, GWs due to deformations on a NS have a frequency of $2 \nu$.

For the \textit{r}-mode scenario, the value for the saturation amplitude that we found to agree well with observation ($\alpha = \num{e-7}$) is many orders of magnitude below what is currently predicted. Theory would need to explain why this is so, or why the instability window is smaller than what is usually assumed.

We have not addressed the spin distribution of RMSPs in this work. Future work could explore how the LMXB population evolves into the RMSP population and consider whether GWs are relevant in this process and can explain the observed distribution.

In our modelling of transient accretion we considered a simple fast-rise, exponential-decay function with a constant average accretion rate. However, in these systems it is expected that binary evolution will play a key role in the accretion rates and result in a long-term modulation of the accretion rate. This, of course, could have a significant effect on the resultant spin distribution. Such long-term variations could be explored in a future study.

\section*{Acknowledgements}

The authors are grateful for useful discussions with B. Haskell and D. I. Jones. NA acknowledges support from the STFC via grant numbers ST/M000931/1 and ST/R00045X/1.




\bibliographystyle{mnras}
\bibliography{bibliography}



%
%


\bsp	
\label{lastpage}
\end{document}